\documentclass[11pt,a4paper]{article}
\usepackage{jcappub}
\usepackage{float}
\usepackage{amssymb}
\usepackage{graphicx}
\usepackage{amsmath}
\usepackage{hyperref}
\usepackage{amsfonts}
\usepackage{subfigure}

\title{Neutron star solutions in perturbative quadratic gravity}

\author[a,b]{Cemsinan Deliduman,}
\author[c]{K. Y. Ek\c{s}i}
\author[c]{and Vildan Kele\c{s}}

\affiliation[a]{Department of Physics, Mimar Sinan Fine Arts University,
Bomonti 34380, \.{I}stanbul, Turkey}
\affiliation[b]{National Astronomical Observatory of Japan, 2-21-1
Mitaka, Tokyo 181-8588, Japan}
\affiliation[c]{\.{I}stanbul Technical University, Faculty of
Science and Letters,\\ Physics Engineering Department,
Maslak 34469, \.{I}stanbul, Turkey}

\emailAdd{cemsinan@msgsu.edu.tr}
\emailAdd{eksi@itu.edu.tr}
\emailAdd{kelesvi@itu.edu.tr}

\abstract{We study the structure of neutron stars in $R+\beta R^{\mu \nu} R_{\mu \nu}$
gravity model with perturbative method.  We obtain
mass--radius relations for six representative equations of state (EoSs). 
We find that, for $|\beta| \sim 10^{11}$ cm$^{2}$, the results differ substantially from the 
results of general relativity. Some of the soft EoSs that are 
excluded within the framework of general relativity
can be reconciled for certain values of $\beta$ of this order 
with the 2 solar mass neutron star recently observed. 
For values of $\beta$ greater than a few 
$10^{11}$ cm$^2$ we find a new solution branch allowing highly massive neutron stars.
By referring some recent observational
constraints on the mass--radius relation we try to constrain the value of $\beta$ for each EoS.
The associated length scale
$\sqrt{\beta}\sim 10^6$ cm is of the order of 
the the typical radius of neutron stars implying that this is the smallest value we could find by using neutron stars as a probe. We thus conclude that the true value of $\beta$ is most likely much smaller than $10^{11}$ cm$^{2}$.}


\begin{document}
\maketitle


\section{Introduction}


Einstein's theory of gravity, general relativity, enjoyed impressive observational and experimental 
support in the past century. Main successes of this theory were explanations it brought to
phenomena such as precession of the perihelion of Mercury, 
bending of light and the gravitational redshift of light near massive bodies. 
Although all of these impressive tests were done solely in the solar system, the 
standard model of cosmology assumes the validity of general relativity 
in all scales up to the very large scale structure of the universe. 
However, as the 20th century was ending, data from
distant supernovae Ia \cite{per99,rie98,rie04} were
interpreted as evidence of late time acceleration in the expansion rate
of the universe. If we continue to assume the validity of general
relativity in all scales, then the best fit to observational data
requires us to introduce a non--vanishing positive cosmological constant into the theory. 
This is the simplest way to follow without changing the basic paradigm. 
Yet there are several theoretical problems related with
the existence of cosmological constant (see for example
\cite{Weinberg,Peebles,Nobbenhuis,Bousso}). One of the most important among these problems
is the lack of a quantum theoretical method to calculate its
inferred value from cosmological data \cite{car01}. There are several proposals in order to avoid 
the problems of cosmological constant with alternative routes of explanations \cite{uzan,tsu10}.
These proposals can be collected into a few groups: to explain late time accelerated expansion 
one can either modify general relativity by modifying the Einstein--Hilbert action, or add new 
gravitational degrees of freedom other than the metric to the theory of gravity, or change how the 
matter fields and perhaps the cosmological constant gravitates \cite{man05,clif11}.

A modified gravity theory, which is proposed to solve late time cosmic acceleration problem, should 
also be able to pass several tests before it can be considered a viable theory of gravity. 
First of all, in the weak gravity regime, such a theory should  be compatible with the solar system
tests and table--top experiments. In cosmological scales, other than
producing the late time accelerated expansion, it should be free of gravitational
instabilities, and obey constraints of the standard model of cosmology.
Such a theory is also expected to do well in strong gravity regime: for
example it should have solutions for neutron stars with mass--radius
relation inside the current observational bounds. There are alternatives to and
generalizations of general relativity theory which have the same predictions in the weak-field
regime as the general relativity, and also provide an explanation of the evolution of the universe in the 
large scale. Thus, the difference between general relativity and alternatives might become prominent 
in the strong-gravity regime \cite{psa08}.

One important family of modifications of Einstein--Hilbert action is
the $f(R)$ theories of gravity (see reviews
\cite{Odintsov-rev,Capozziello-rev,Sotiriou-rev,deFelice-rev}
and references therein). In such theories one uses a function of curvature scalar as the
Lagrangian density. The $f(R)$ term must have a lower order expansion in Ricci scalar in order
to include general relativity as, perhaps, a weak-field limit of it. Such models of gravity
can be made to pass Solar System tests, explain the late-time accelerated
expansion of the universe and also work well in the strong-field regime.
In a previous work \cite{ara11} a simple version of $f(R)$ gravity theory, 
in which $f(R) = R +\alpha R^2$, is studied by two of the present authors. 
It is curious that predictions of such theories can be shown to be 
equivalent to scalar-tensor theories of gravity
\cite{mag94}, which have been analyzed throughly since the seminal paper of Brans and Dicke 
\cite{bd}. However, if we modify the Einstein--Hilbert term 
by contractions of Ricci and Riemann tensors, 
$R_{\mu \nu }R^{\mu \nu }$ and $R_{\mu \nu \rho \sigma}R^{\mu \nu \rho \sigma}$,
then we have an alternative gravity theory with independent predictions.

Inspiration for such an alternative gravity theory may come from string theory. Absence of ghosts in low 
energy string theory in flat backgrounds requires the quadratic corrections to Einstein's gravity to be of 
the Gauss-Bonnet (GB) form \cite{zwi85}:
\begin{equation}
S=\int d^{4}x\sqrt{-g}\left[ R+\gamma ( R^2 -4 R_{\mu \nu }R^{\mu \nu }
+R_{\mu \nu \rho \sigma}R^{\mu \nu \rho \sigma} )\right]\ .
\label{zwei}
\end{equation}
However, the Gauss--Bonnet term will not contribute to the equations of motion, because it is 
equivalent to a total derivative in four dimensions. This means that variations of 
$R_{\mu \nu \rho \sigma}R^{\mu \nu \rho \sigma}$ can be expressed in terms of variations of 
$R^2$ and $R_{\mu \nu }R^{\mu \nu }$. Contraction of Riemann tensors,
could have relevance only in the cases related to quantum gravity \cite{psa09,par99}. 
Therefore for the classical physics applications one can 
take the action of our string-inspired gravity as 
\begin{equation}
S=\int d^{4}x\sqrt{-g}\left[ R+\alpha R^2 +\beta R_{\mu \nu }R^{\mu \nu }\right]\ ,
\label{rqugr}
\end{equation}
where $\alpha $ and $\beta $ are free-parameters, taken independent of each other. 
However, special relations between $\alpha $ and $\beta $ exists and such cases
correspond to some special theories: 
for example, $\beta=-3\alpha$ correspond to the Weyl tensor squared modification of general relativity 
\cite{s08,fs09}, and $\beta=-4\alpha$ has unique energy properties \cite{bay02}. 

As mentioned above, the absence of ghosts in the low energy string theory in flat backgrounds requires 
the quadratic corrections to Einstein's gravity to be of the form given in equation (\ref{zwei}). This 
means that the alternative gravity theory defined by (\ref{rqugr}) cannot be free of ghosts and this 
brings up the question of stability of the neutron star solutions discussed in this paper. However, all 
such modified gravity theories, bar the theory defined with (\ref{zwei}), suffer from the same problem, 
the solution of which is beyond the scope of this paper. Our aim in this paper is to see the possibility
of neutron star solutions in a specific modified gravity theory defined by (\ref{rqugr}) with $\alpha =0$,  
to analyze solutions with mass-radius relations obeying the observational constraints, and to see 
if there are problems in this gravity model other than the well known stability issue.
When combined with the results of \cite{ara11} this analysis will allow for a comparison of the effects of $R^2$ and $R_{\mu \nu }R^{\mu \nu }$ terms in (\ref{rqugr}) on the structure of neutron stars.

We further note that the mass dimensions of quadratic corrections to 
Einstein--Hilbert action in (\ref{rqugr}) is 
order $[L]^{-4}$. It is possible to add terms with the same mass dimensions to the above action. In fact 
referring to the quantum gravity arguments in \cite{par99}, we could also add the following dimension
$[L]^{-4}$ terms to the action:
\begin{equation}
\square R\quad \mathrm{and}\quad \nabla ^{\mu }\nabla ^{\nu }R_{\mu \nu }\ .
\end{equation}
However, as it is pointed out in \cite{emi10} these terms do not contribute to the field equations and 
therefore in the context of the present paper they are also redundant.

The effect of the value of $\alpha $, while $\beta =0$, on the mass-radius (M-R) relation of
neutron stars has been studied, for a polytropic EoS in \cite{coo10}, 
and for realistic EoS in \cite{ara11}. 
The latter work constrained the value of $\alpha $ to be $|\alpha |\lesssim 10^{10}$ cm$^{2}$.
In an other study, Santos analyzes neutron stars in this model of gravity 
for a single EoS of ideal neutron gas \cite{emi11}
and for a specific choice of $\beta=-2\alpha$, $\sqrt{\alpha}=0.96$ km.
In that work, the author shows, for this restricted choice of EoS and parameters, that
stable configurations of neutron stars are possible even for arbitrarily large baryon numbers
of the neutron star.

Our approach in this work is different than \cite{emi11} in two ways: (i) Rather than choosing
a fixed specific value for $\beta$, while $\alpha =0$, we study its effect, as a free parameter, 
on the M-R relation and constrain its
value by referring to observations, (ii) As the interaction between 
nucleons can not be neglected, the EoS of ideal neutron gas
can not provide realistic M-R relations. 
We thus employ six different realistic EoSs corresponding to a variety of
assumptions for the interaction between nucleons rather
than the very restrictive EoS of ideal neutron gas.

In the present work, we adopt an alternative theory of gravity, in which the Einstein-Hilbert
action is modified with the term $R_{\mu\nu}R^{\mu\nu}$ only.  
In \S 2, we assume a perturbative form of our alternative gravity model and obtain the
field equations. The reason of perturbative approach is that the equations of motion derived
from this alternative gravity theory are fourth order differential equations, 
and their treatment in four dimensions is problematic. 
To obtain the modified Tolman--Oppenheimer--Volkoff (TOV) equations from the field equations
we also assumed perturbative forms of metric and hydrodynamical functions.
In \S 3, we solve the structure of neutron stars in this gravity model for six
representative equations of state describing the dense matter of neutron
stars. We plot the mass-radius relations of neutron stars for $\beta$ changing in the range 
$\sim \pm 10^{11}$ cm$^2$. 
This way the value of the perturbative parameter $\beta$ is constrained by the recent 
measurements of the mass-radius relation \cite{oze10} and the observed 2 solar mass neutron star \cite{dem10}. 
We identify that $\beta \sim \pm 10^{11}$ cm$^2$ produces results that can have observational
consequences. Lastly, the results of the numerical study and the significance of the scale
of the perturbation parameter is discussed in the conclusions.


\section{Modified TOV equations}


The approach that we are going to take in this paper is in the 
spirit of \cite{ara11}, although not in details. 
The alternative theory that we are going to analyze is not the full quadratic gravity, but part of it whose 
analysis would complement the analysis reported in \cite{ara11}. In the appropriate units, the defining 
action of our alternative theory is given as 
\begin{equation}
S=\int d^{4}x\sqrt{-g}\left( R+\beta R_{\mu \nu }R^{\mu \nu }\right)
+ S_{{\rm matter}}\; ,
\label{lagrangian}
\end{equation}
where we already set $G=1$ and $c=1$. This will also be the case in
the rest of this section. We note that, in this study we are using the metric formalism of gravity in which 
matter only couples to the metric, and the Levi--Civita connection is a function of the metric.
Variation of the action (\ref{lagrangian}) with
respect to the metric results in the field equations,
\begin{eqnarray}
8\pi T_{\mu \nu } &=&G_{\mu \nu }+\beta \left( -\frac{1}{2}g_{\mu \nu
}R_{ab}R^{ab}+\nabla ^{\rho }\nabla _{\rho }R_{\mu \nu }\right)  \notag \\
&&+\beta \left( -\nabla _{\nu }\nabla _{\mu }R-2R{}_{\sigma \mu \nu \alpha
}R^{\alpha \sigma }+\frac{1}{2}\square Rg_{\mu \nu }\right).  \label{EoM}
\end{eqnarray}
where $G_{\mu \nu}=R_{\mu \nu }-\frac{1}{2}Rg_{\mu \nu }$ is the Einstein tensor.
The energy--momentum tensor, $T_{\mu \nu }$, is the energy-momentum tensor of the perfect fluid. 

In the second step we derive the hydrostatic equilibrium equations within the
framework of the gravity model considered. The hydrostatic equilibrium
equations, obtained and solved by Tolman-Oppenheimer and Volkoff \cite {tol39,opp39} within the 
framework of general relativity, are commonly called TOV equations. We use the same 
nomenclature in this paper, although the hydrostatic equilibrium equations in this gravity 
model will turn out to be quite different.

As in the case of general relativity we choose to work with a spherically symmetric
diagonal form of the metric and metric functions depend only on the radial coordinate $r$:
\begin{equation}
ds^{2}=-e^{2\phi_{\beta }}dt^{2}+e^{2\lambda_{\beta }}dr^{2}+r^{2}
\left(d\theta^{2}+\sin ^{2}\theta d\phi ^{2} \right) \label{metric}
\end{equation}

When this metric ansatz and the energy--momentum tensor for perfect fluid substituted into the field 
equations (\ref{EoM}), the resulting equations will contain metric functions $\phi (r)$ and $\lambda (r)$, 
hydrodynamic quantities $P(r)$ and $\rho(r)$, as well as their derivatives with respect to 
$r$. The presence of these higher order derivatives precludes expressing the equation 
in terms of hydrodynamic quantities only and consequently deriving the modified TOV equations. 
In order to reach that aim we are going to use the
perturbative approach \cite{ja86, eli89} where general relativistic solution is taken as the 
zeroth order solution of the field equations (\ref{EoM}).
As mentioned, this method had already been applied to $f(R)$ models of gravity
via perturbative constraints at cosmological scales \cite{ded08, coo09} and
neutron stars with $f(R)=R + \alpha R^{n+1}$ \cite{coo10, ara11}. In the
perturbative approach, $g_{\mu \nu }$ can be expanded perturbatively in
terms of $\beta $: 
\begin{equation}
g_{\mu \nu} = g_{\mu \nu}^{(0)} + \beta g_{\mu \nu}^{(1)} + O(\beta^2)
\end{equation}
Accordingly, the metric functions must also be expanded in terms of $\beta $ as 
\begin{equation}
\phi _{\beta }=\phi +\beta \phi_{1}+\cdots \quad \mathrm{and}\quad 
\lambda _{\beta }=\lambda +\beta \lambda_{1}+\cdots ,
\end{equation}
and the hydrodynamic quantities of the perfect fluid are expanded perturbatively as:
\begin{equation}
\rho _{\beta }=\rho +\beta \rho_{1}+\cdots \quad \mathrm{and}\quad
P_{\beta }=P+\beta P_{1}+\cdots .
\end{equation}
In these equations functions without subscript $\beta$ are general relativistic solutions. In the present 
perturbative approach they are ``zeroth order'' solutions and obtained from Einstein equations,
\begin{eqnarray}
  -8\pi \rho &=& -r^{-2} +e^{-2\lambda}(1-2r\lambda')r^{-2} \label{e-tt},\\
  8\pi P &=& -r^{-2} +e^{-2\lambda}(1+2r\phi')r^{-2}, \label{e-rr}
\end{eqnarray}

In order to determine the first and the second TOV equations, 
we now evaluate $tt$ and $rr$ components of field equations (\ref{EoM}) in terms of metric functions
(\ref{metric}) and hydrodynamic quantities $P(r)$ and $\rho (r)$. The $tt$ component of the field 
equations turns out to be
\begin{eqnarray}
8\pi \rho _{\beta } &=&\frac{1}{r^{2}}e^{-2\lambda _{\beta }}\left(
2r\lambda _{\beta }^{\prime }-1+e^{2\lambda _{\beta }}\right) 
+\frac{1}{2} \beta \left( \rho ^{2}+3P^{2}\right)  \notag \\
&&+\frac12 \beta e^{-2\lambda} \left[
\begin{array}{l}
\smallskip
\rho ^{\prime \prime}+3P^{\prime \prime }
- R^{\prime \prime }+\left( 2\lambda ^{\prime }+ 2\phi ^{\prime } 
- \frac{1}{2}\frac{g^{\prime }}{g}\right) R^{\prime } \\
\smallskip
+ \left( \phi ^{\prime }-\lambda ^{\prime } +\frac2r \right) ( \rho ^{\prime }+3P^{\prime}) 
-4(\phi ^{\prime })^{2}\left( \rho +P\right) \\
\smallskip
-2 \left( \phi ^{\prime }\lambda ^{\prime }-( \phi ^{\prime }) ^{2}-\phi ^{\prime \prime } 
-\frac2r \phi ^{\prime }\right) ( \rho -P)  
\end{array}
\right] ,  \label{dilaver}
\end{eqnarray}
and the $rr$ component of the field equations is
\begin{eqnarray}
8\pi P_{\beta } &=&\frac{1}{r^{2}}e^{-2\lambda _{\beta }}\left( 2r\phi _{\beta }^{\prime
}+1-e^{2\lambda _{\beta }}\right) -\frac{1}{2}\beta
\left( \rho ^{2}+3P^{2}\right)  \notag \\
&&+\beta e^{-2\lambda }\left[ 
\begin{array}{l}
\smallskip
P^{\prime \prime } +\left( \phi ^{\prime } +\frac2r \right) \rho ^{\prime }
-\left( 2\phi ^{\prime } +\lambda ^{\prime } +\frac4r \right) P^{\prime } \\
\smallskip
+ \left( \phi ^{\prime \prime } - \phi ^{\prime }\lambda ^{\prime } 
-( \phi ^{\prime }) ^{2} +\frac2r \lambda ^{\prime } \right) \rho \\
\smallskip
+ \left( 3\phi ^{\prime \prime } - 3\phi ^{\prime }\lambda ^{\prime } 
+( \phi ^{\prime }) ^{2} -\frac2r \lambda ^{\prime } \right) P
\end{array}
\right] .  \label{veli}
\end{eqnarray}
Here, prime denotes derivative with respect to $r$. 
Note that since terms multiplied with $\beta$ are already first
order in $\beta$, metric and hydrodynamic functions that appear in those terms are taken as zeroth 
order general relativistic functions. In all the other terms we have full functions. However, when we 
integrate TOV equations numerically, we are going to keep terms only up to first order in $\beta$.

We now define a mass parameter $m_{\beta}$ by the relation $e^{-2\lambda _{\beta }}=1-m_{\beta }/r$
so that the solution for the metric function $\lambda _{\beta } (r)$ would have the same form of the 
exterior solution. Here, again we take $m_{\beta }$ to be expanded in $\beta $ as $m_ {\beta }=m+
\beta m+ \cdots$ where $m$ is the general relativistic zeroth order solution. It is given in terms of 
$\rho (r)$ as
\begin{equation}
m=8\pi \int \rho \left(r \right) r^{2} dr
\end{equation}
Taking the derivative of mass parameter, $m_{\beta}$, with respect to $r$ we obtain
\begin{equation}
\frac{dm_{\beta }}{dr}=e^{-2\lambda _{\beta }}( 2r\lambda _{\beta
}^{\prime }-1+e^{2\lambda _{\beta }}) .  \label{deneb}
\end{equation}

In the perturbative approach all terms which are multiplied with $\beta$ in (\ref{dilaver}, \ref{veli}) 
can be rearranged in terms of  zeroth order general relativistic expressions by ignoring 
higher order terms. Using the expression (\ref{deneb}) for $\dfrac{dm_{\beta }}{dr}$ in $tt$ field 
equation (\ref{dilaver}) we obtain the first modified TOV equation as
\begin{eqnarray}
\frac{dm_{\beta }}{dr} &=&8\pi \rho _{\beta }r^{2}
-\beta r^{2} \left[
\begin{array}{l}
\smallskip
3\left( 1-\frac{m}{r}\right) P^{\prime \prime}
- \frac32 \left( \rho r+3\frac{m}{r^{2}}-\frac{4}{r}\right) P^{\prime }  
+ \frac12 \left( Pr+\frac{m}{r^{2}}\right) \rho ^{\prime} \\
+ \rho ^{2}+P\rho
- \frac{\rho +P}{2(r-m)}\left( P^{2}r^{3}+\frac{m^{2}}{r^{3}}+2Pm\right)
\end{array}
\right] .  \label{first_tov}
\end{eqnarray}

To derive the second modified TOV equation we use 
the continuity equation of energy-momentum tensor, $\nabla^\mu T_{\mu\nu}=0$, which is 
equivalent to the hydrostatic equilibrium condition,
\begin{equation}\label{hydro}
    \frac{dP_\beta}{dr}=-(\rho_\beta
    +P_\beta)\frac{d\phi_\beta}{dr}.
\end{equation}
We, therefore, need to read $\dfrac{d\phi_\beta}{dr}$ from the $rr$ field equation (\ref{veli})
and use in the above expression to obtain the second modified TOV equation.
After some straightforward algebra we find it as
\begin{equation}
2\left( r-m_{\beta }\right)\frac{d\phi_\beta}{dr} =
8\pi r^{2}P_{\beta }+\frac{m_{\beta }}{r} -\beta r^{2}\left[ 
\begin{array}{l}
\smallskip
\left( 1-\frac{m}{r}\right) P^{\prime \prime }  
- \left( Pr+\frac{1}{2}\rho r-\frac{7}{2}\frac{m}{r^{2}}+\frac{4}{r}\right) P^{\prime } \\
\smallskip
+ \left( \frac{1}{2}Pr+\frac{2}{r}-\frac{3}{2}\frac{m}{r^{2}}\right) \rho ^{\prime } \\
\smallskip
\rho ^2+P\rho -2\frac{m}{r^{3}}\left( \rho +P\right) \\
- \frac{\rho +P}{2(r-m)}\left( P^{2}r^{3}+\frac{m^{2}}{r^{3}}+2Pm\right) 
\end{array}
\right] .  \label{emzade}
\end{equation}

Note that one gets the original TOV equations in general relativity when one sets $\beta$ 
to zero and transforms $m\rightarrow 2m$. The modified TOV equations, (\ref{first_tov}),
(\ref{hydro}) and (\ref{emzade}), are complicated and therefore similar to the case 
in general relativity they are solved numerically. 
We explain the  numerical analysis and present the results of it in the next section.


\section{Solution of the neutron star structure}


In this section we numerically solve the modified TOV equations, namely
eqs. (\ref{first_tov}) and (\ref{hydro}), with realistic equations of state (EoSs) appropriate for
neutron stars.

\subsection{The Equations Solved}

With the physical constants plugged and $m\rightarrow 2m$ transformed, 
Eqns.(\ref{first_tov}) and (\ref{hydro})
become 
\begin{equation}
\frac{dm}{dr}=4\pi r^{2}\rho +\frac{1}{2}\beta r^{2}K  \label{dmdr}
\end{equation}%
where 
\begin{eqnarray}
K &=&-\left( 1+\frac{P}{\rho c^{2}}\right) \frac{G\rho ^{2}}{c^{2}} 
-\left( 1-\frac{2Gm}{c^{2}r}\right) 3\frac{P^{\prime \prime }}{c^{2}}
-\left( 1+\frac{2mc^{2}}{r^{3}P}\right) \frac{GrP\rho ^{\prime }}{2c^{4}} 
\nonumber \\
&&+\frac{Gm}{rc^{2}}\frac{GP\rho }{2c^{4}}\left( 1+\frac{P}{\rho c^{2}}%
\right) \left( \frac{Pr^{3}}{mc^{2}}+\frac{4mc^{2}}{r^{3}P}+4\right) \left(
1-\frac{2Gm}{rc^{2}}\right) ^{-1}  \nonumber \\
&&+\left( \frac{\rho c^{2}}{P}+\frac{6mc^{2}}{r^{3}P}-\frac{4c^{4}}{Gr^{2}P}%
\right) \frac{3Gr^{2}PP^{\prime }}{2rc^{6}}  \label{K}
\end{eqnarray}%
and 
\begin{equation}
\frac{dP}{dr}=-\frac{Gm\rho \left( 1+\frac{P}{\rho c^{2}}\right) }{%
r^{2}\left( 1-\frac{2Gm}{c^{2}r}\right) }\left( 1+\frac{4\pi r^{3}P}{mc^{2}}+%
\frac{1}{2}\beta r^{2}H\right)   \label{dPdr}
\end{equation}%
where 
\begin{eqnarray}
H=&&-\left(1-\frac{2Gm}{rc^{2}}\right)\frac{rP^{\prime \prime }}{mc^{2}}  
-\left(\frac{1}{2}+\frac{2c^{4}}{Gr^{2}P}-\frac{3mc^{2}}{r^{3}P}\right)\frac{%
Gr^{2}P\rho ^{\prime }}{mc^{4}}  \nonumber \\
&&-\left( \frac12 \frac{\rho ^{2}c^{4}}{P^{2}}+\frac{1}{2}\frac{\rho
c^{2}}{P}-\frac{2mc^{2}}{r^{3}P}\frac{\rho c^{2}}{P}-\frac{2mc^{2}}{r^{3}P}%
\right) \frac{2GrP^{2}}{mc^{6}}  \nonumber \\
&&+\left( 1+\frac{1}{2}\frac{\rho c^{2}}{P}-\frac{7mc^{2}}{r^{3}P}+\frac{%
4c^{4}}{Gr^{2}P}\right) \frac{GP^{\prime }r^{2}P}{mc^{6}}  \nonumber \\
&&+\left(\frac{1}{4}\frac{r^{3}P}{mc^{2}}+\frac{mc^{2}}{r^{3}P}+1 \right)\left( 1-%
\frac{2Gm}{rc^{2}}\right)^{-1}\left( 1+\frac{P}{\rho c^{2}}\right) \frac{P\rho 
}{c^{2}}\frac{2mG^{2}}{mc^{4}}.  \label{H}
\end{eqnarray}

The dimension of the coupling constant $\beta $ is length square. Therefore,
perturbation is actually over the dimensionless quantity $\bar{\beta}=\beta
/\beta _{0}$, where $\beta _{0}$ is the nominal value appropriate for neutron stars.
It is possible to define it in terms of fundamental constants as follows.
The typical mass of degenerate stars, including neutron stars, in terms of fundamental constants is
\begin{equation}
M_{0}=\frac{m_{\mathrm{Pl}}^{3}}{m_{\mathrm{n}}^{2}}=1.8482M_{\odot }
\end{equation}%
where $m_{\mathrm{Pl}}$ is the Planck mass and $m_{\mathrm{n}}$ is the nucleon
mass. Note that this can also be written as $M_{0}=\alpha_{\mathrm G}^{-3/2}m_{\mathrm{n}}$ where $\alpha_{\mathrm G}=Gm_{\mathrm{n}}^2/\hbar c$ is the ``gravitational coupling constant.'' The radius of the neutron star which is about 10-15 km can be given in terms of this mass as three times its Schwarzschild radius. As $\beta_0$ has the dimension of length square, it will be given as the square of this typical radius which is the only length scale in the system. Thus
\begin{equation}
\beta _{0}=\left( \frac{6GM_{0}}{c^{2}}\right) ^{2}=268.128\, \mathrm{km}^{2}\simeq 2.7\times 10^{12}\, \mathrm{cm}^{2}.
\end{equation}%
Accordingly,
we can expect that deviations from general relativity (GR), which come through the parameter 
$\beta $, become significant when $\beta $ is a sizeable fraction of $\beta
_{0}$. This is indeed what we observe through the numerical solution of the
equations.

Another way to estimate the order of magnitude  of the coupling constant $\beta$ is to see that 
a characteristic value of the Ricci scalar, and the square root of $R^{\mu \nu}R_{\mu \nu}$, 
is given by $R_0 \sim GM_{\ast}/c^2 R_{\ast}^3 \sim G \rho_c / c^2$. 
As $\beta_0 \sim 1/R_0$ we infer that $\beta_0 \sim c^2/G\rho_c$. In Section  3.6 we check the 
validity of the perturbative approach by plotting the dimensionless coupling constant $\beta/\beta_0$ for a range of static mass configurations.

In the above equations, the higher derivatives, like $P^{\prime }$, $\rho
^{\prime }$ and $P^{\prime \prime }$ are calculated by using the TOV
equations obtained in GR. As such terms come only in the perturbative term,
the error in employing GR instead of the full theory is of the order of $%
\beta ^{2}$.

\subsection{Equation of State}

In order to solve the hydrostatic equilibrium equations (\ref{dmdr}) and (%
\ref{dPdr}) we must supplement them with an equation of state (EoS) defining
the microscopic physics of neutron stars (NSs). The EoS of NS matter at the inner core where
most of the mass resides is not well constrained. Different EoSs lead to
different mass-radius (M-R) relations. As a result we have to solve the NS
structure for a number of EoSs in order to show that our basic result, that $%
|\beta |\lesssim 10^{12}$ cm$^{2}$, does not depend on the EoSs. In this work
we present results for six representative EoSs including typical EoS involving only nucleons as well as 
EoS involving nucleons and exotic matter. We do not make any analysis on strange quark stars as 
these are not gravitationally bound objects. We use an analytical
representation of $\log \rho(\log P)$ for all the EoSs obtained by fitting
the tabulated EoS data following the method described in \cite{ge11}. These EoSs are 
FPS \cite{ref_FPS}, AP4 \cite{ref_AP4}, SLy \cite%
{ref_SLY}, MS1 \cite{ref_MS1}, MPA1 \cite{ref_MPA1} and GS1 \cite{ref_GS1}.
The physical assumptions of these EoSs are described in \cite{lat01}.

\subsection{Numerical Method}

We employ a Runge-Kutta scheme with fixed step 
size of $\Delta r=0.01$ km starting from the center of the star for a certain value
of central pressure, $P_{\mathrm{c}}$. We identify the surface of the
star as the point where pressure drops to a very small value 
(10 dyne/cm$^{2}$) and record the mass contained inside this radius as the mass of the
star.

We change the central pressure $P_{\mathrm{c}}$ from $3\times 10^{33}$ dyne
cm$^{-2}$ to $9\times 10^{36}$ dyne cm$^{-2}$ in 200 logarithmically equal
steps to obtain a sequence of equilibrium configurations. We record the mass
and radius for each central pressure. This allows us to obtain a mass-radius
(M-R) relation for a certain EoS. We then repeat this procedure for a range
of $\beta $ to see the effect of the perturbative term we added to the
Lagrangian.

\subsection{Observational Constraints on the Mass-Radius Relation}

In order to constrain the value of $\beta$ we have used the recent measurements
of mass and radius of neutron stars, EXO 1745-248 \cite{oze1745}, 4U 1608-52 \cite{guv1608} and 4U 1820-30 \cite%
{guv1820} (see \cite{oze09} for a description of the method used). 
We use the
$2\sigma$ confidence contours, shown in all M-R plots as the region bounded by the thin black
line, on the M-R relation of neutron stars given in \cite{oze10},
which is a union of these three constraints.\footnote{%
see \cite{ste10} for a critic of these constraints .} 

Apart from the above we also use the 
mass of PSR J1614-2230 with $1.97\pm 0.04\,M_{\odot }$ as recently measured by \cite{dem10},
as a constraint. It is shown as the
horizontal black line with grey error-bar. For a viable combination
of $\beta $ and EoS, the maximum mass of the neutron star must exceed
this measured mass.

These two constraints we employed exclude many of the possible EoSs within the framework of GR. 
The value of $\beta$, the coupling parameter of the gravity model studied in the present work, 
is a new degree of freedom which can alter the M-R relation if it takes values of order $10^{11}$ cm$^2$.
By using this freedom one can determine the range of $\beta$ values for each EoS that is consistent with the observational constraints. 

\subsection{The effect of $\beta $ on the M-R relation}

We have determined the M-R relation for each EoSs for a range of $\beta$
values. Results for each six representative EoSs are summarized below.
In the following we use $\beta _{11}\equiv \beta /10^{11}$ cm$%
^{2}$ instead of $\beta$ as we are mostly interested with $\beta$ of this order. 
For all EoS we observe that the maximum mass of a NS
increases with decreasing value of $\beta_{11}$ while the radius becomes smaller.

\subsubsection{FPS}
\label{sub_sub_FPS}

\begin{figure}
\centering
\mbox{
\subfigure[]{
\includegraphics[width=0.45\textwidth]{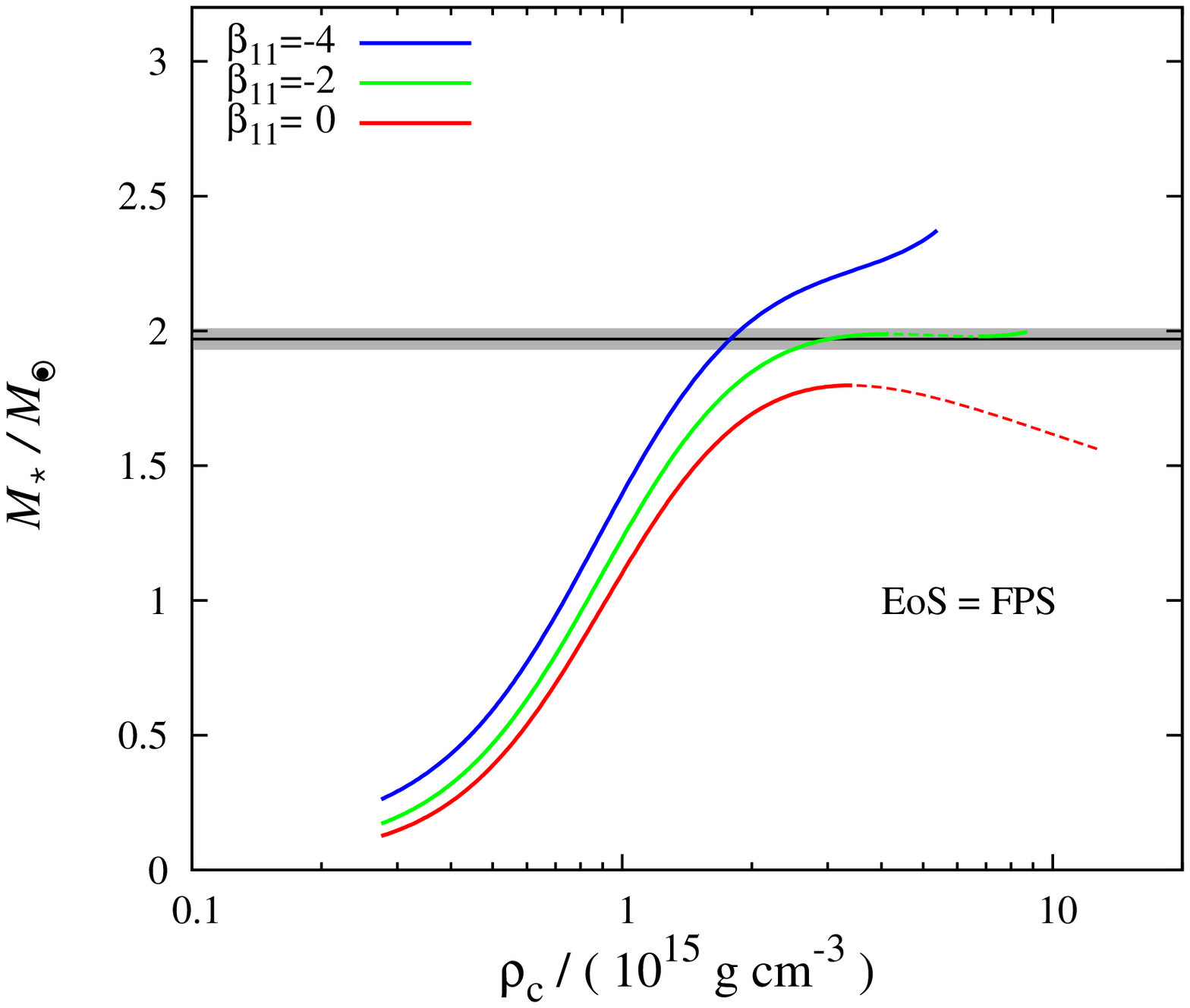}
\label{FPS_rho_M}
}\quad
\subfigure[]{
\includegraphics[width=0.45\textwidth]{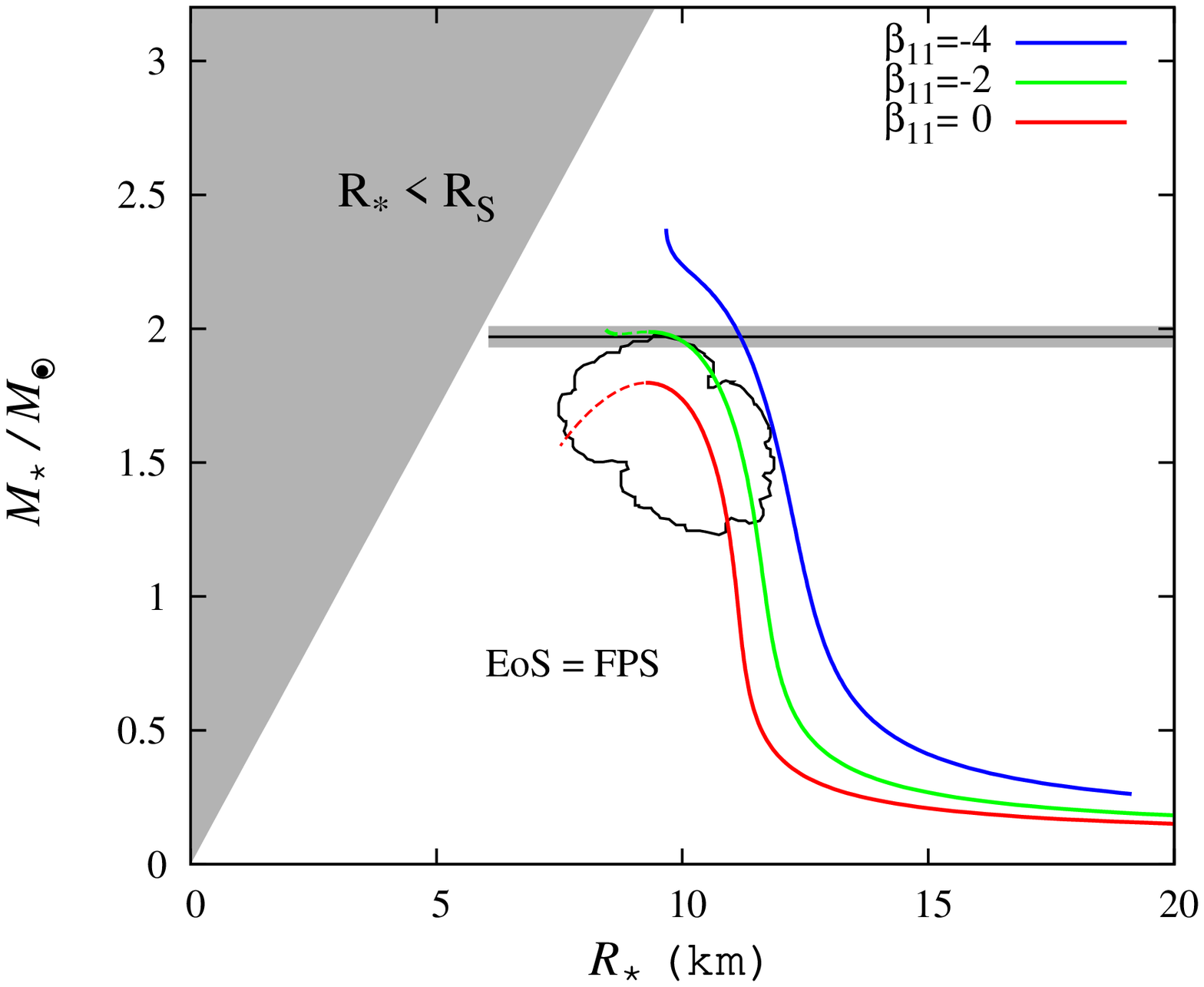}
\label{FPS_MR}
}
}
\centering
\caption{The $\rho_\mathrm{c}-M$ (panel a) and M-R (panel b) relation for FPS. 
The solid lines
correspond to stable configurations for different values of $\beta$. 
The dashed lines correspond to the unstable configurations for which $dM/d\rho_\mathrm{c} < 0$. 
The red line ($\beta=0$) stands for the results in GR.
The thin black line on the right panel shows the $2\sigma$ confidence 
contour of the observational measurements 
of \cite{oze10}; the measured mass $M=1.97 \pm
0.04\, M_{\odot}$ of PSR J1614-2230 \cite{dem10} is shown as the
horizontal black line with grey error bar. The grey shaded region 
shows where the radius of the NS
would be less than the Schwarzschild radius. We observe that
$M_{\max}$ and $R_{\min}$ increase for decreasing values of $\beta$. 
The comparison with observational constraints on the Figure (see the discussion in \S \ref{sub_sub_FPS}) 
imply that $-4<\beta_{11}<-2$ 
in this gravity model if FPS is assumed to be the EoS of NSs.
}
\label{fig-FPS}
\end{figure}

The relation between the central density and the mass of the neutron star, $\rho _{%
\mathrm{c}}-M$, and the M-R relation are shown in Figure~\ref{fig-FPS}. 
For FPS, the maximum mass within GR, is about $1.8\,M_{\odot }$ which is less 
than the $2M_{\odot}$ of PSR J1614-2230  meaning that FPS is excluded within GR ($\beta =0$).

From Figure~\ref{FPS_MR} we  find that, for $\beta
_{11}=-2$, the maximum mass becomes $M_{\max }\simeq 2M_{\odot }$. Thus
FPS can be reconciled with the maximum mass constraint for 
$\beta_{11} <-2$.

For values of $\beta_{11}<-4$ we see that the M-R relation does not pass through the confidence 
contours of the
measured \cite{oze10} mass and radius. This then implies that $\beta_{11}>-4$. Together with the 
previous constraint we conclude that FPS is consistent with observations if $-4<\beta_{11}<-2$ in this 
gravity model.

Interestingly, we find that for $\beta_{11} <-2$ there
exists a new branch of stable solution, in the sense that $dM/d\rho_c>0$, at highest densities which 
does not have a counterpart in GR.

For $\beta_{11}=-3$ and lower values we find that, for the highest central densities, $dm/dr<0$ which 
indicates an unstable star. As a result such solutions are excluded from the Figures~\ref{FPS_rho_M} 
and \ref{FPS_MR}.

\subsubsection{SLy}
\label{sub_sub_SLy}

\begin{figure}
\centering
\mbox{
\subfigure[]{
\includegraphics[width=0.45\textwidth]{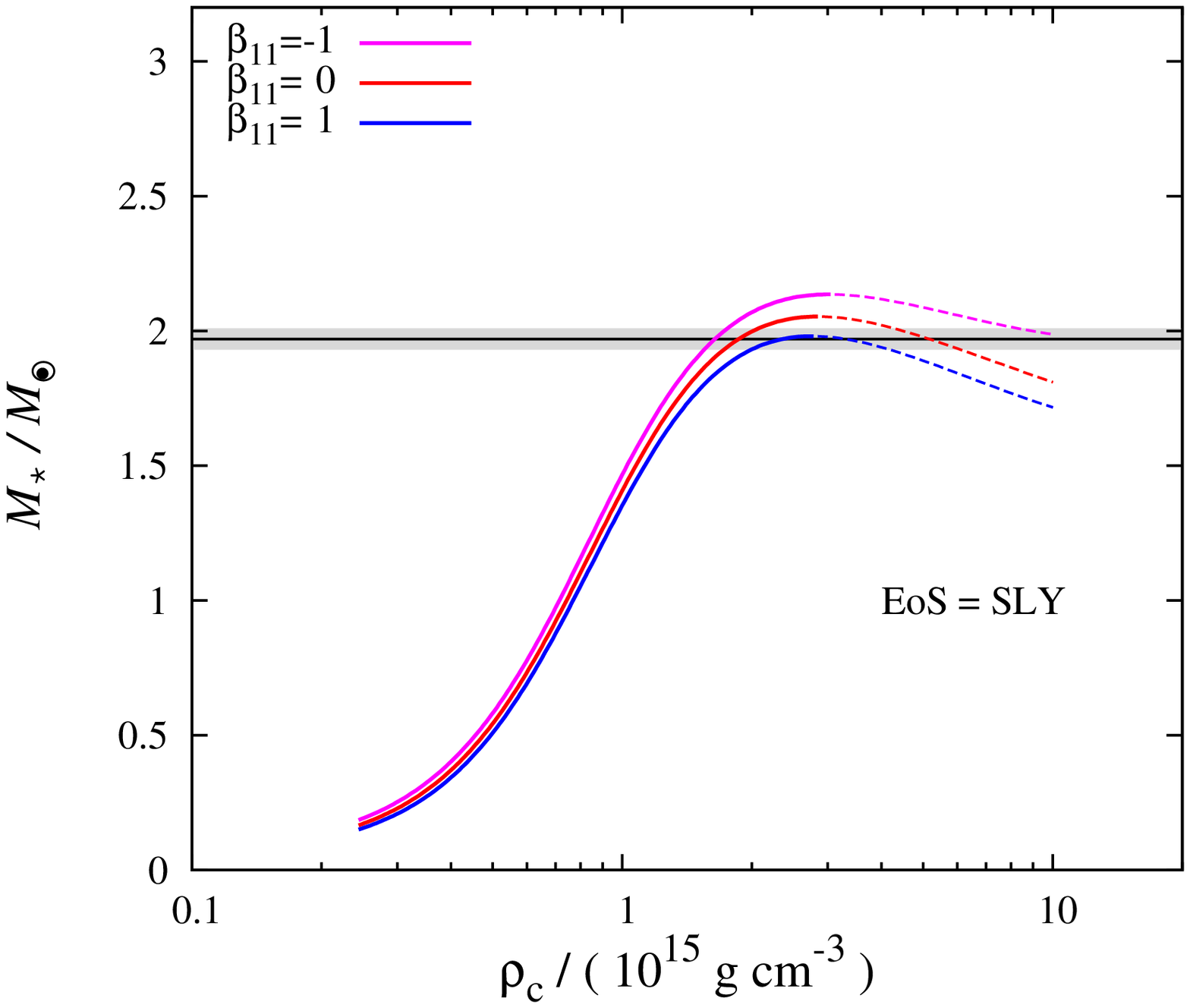}
\label{SLY_rho_M}
}\quad
\subfigure[]{
\includegraphics[width=0.45\textwidth]{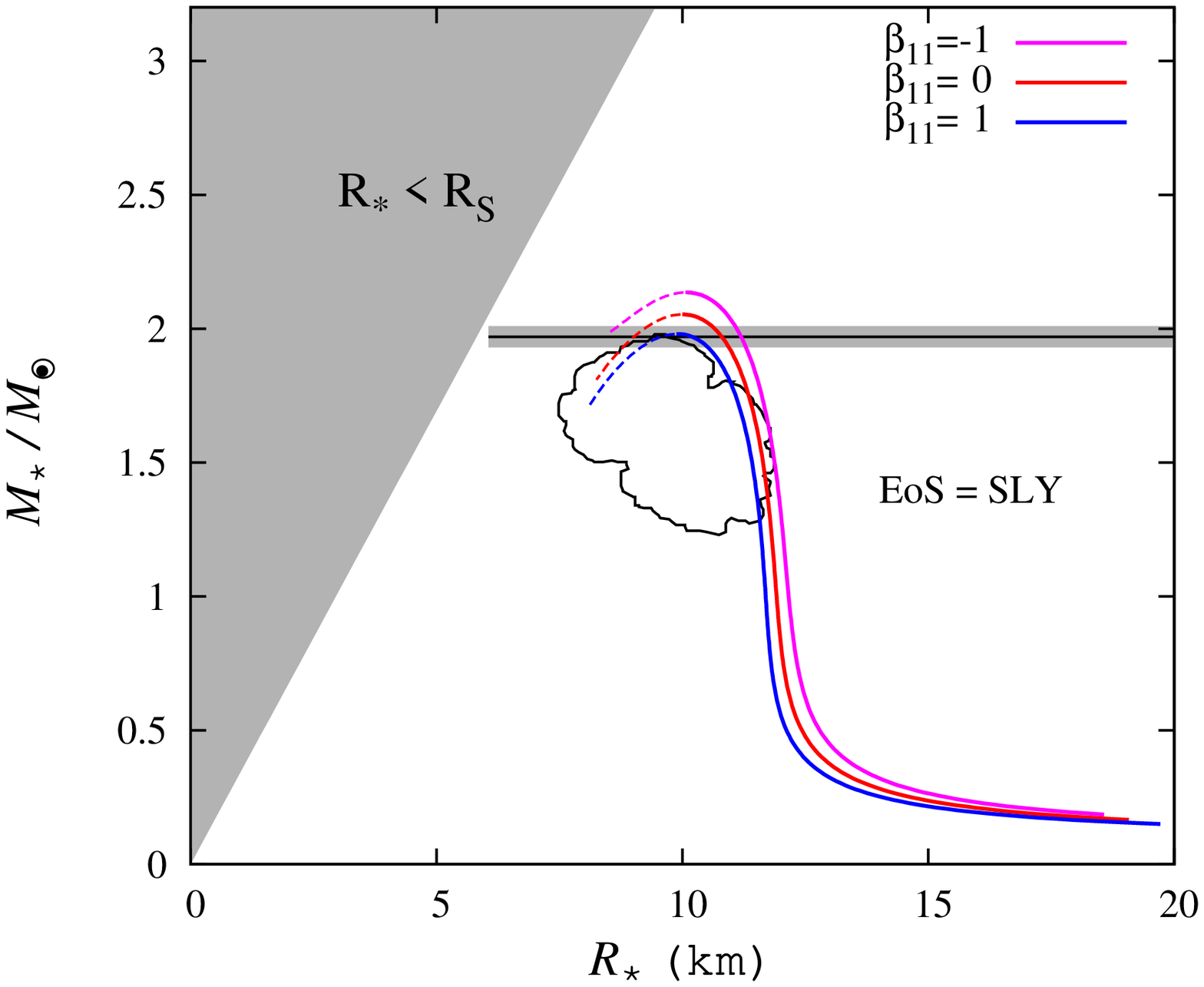}
\label{SLY_MR}
}
}
\centering
\caption{The $\rho_\mathrm{c}-M$ (panel a) and M-R relation (panel b) for SLy EoS. 
See Figure~\ref{fig-FPS}
for the notation
in the figure.
The results
are discussed in \S \ref{sub_sub_SLy}.
}
\label{fig-SLY}
\end{figure}

For SLY the $\rho _{\mathrm{c}}-M$ and M-R relation are shown in Figure~\ref{fig-SLY}.
For $\beta_{11}>2$ we see that $M_{\max}$ is less
than the measured mass of PSR J1614-2230 \cite{dem10} and so is not compatible with its existence. 
For $\beta_{11}<-1$, the M-R relation does not pass through the confidence contours of \cite{oze10}.
These two constraints then imply that $2>\beta_{11}>-1$ for the gravity model employed here so that 
SLy is consistent with the observations.

Again, we see that for $\beta_{11}=-2$ (a value which we have already excluded 
by confronting with observations) a new stable solution
branch exists for highest densities. For even lower values of $\beta$  
we obtain unstable ($dm/dr<0$) solutions for the highest densities. This is why we can not extend
the solution to $\beta_{11}=-3$ for densities greater than $5 \times 10^{15}$ g cm$^{-3}$.

\subsubsection{AP4}
\label{sub_sub_AP4}

\begin{figure}
\centering
\mbox{
\subfigure[]{
\includegraphics[width=0.45\textwidth]{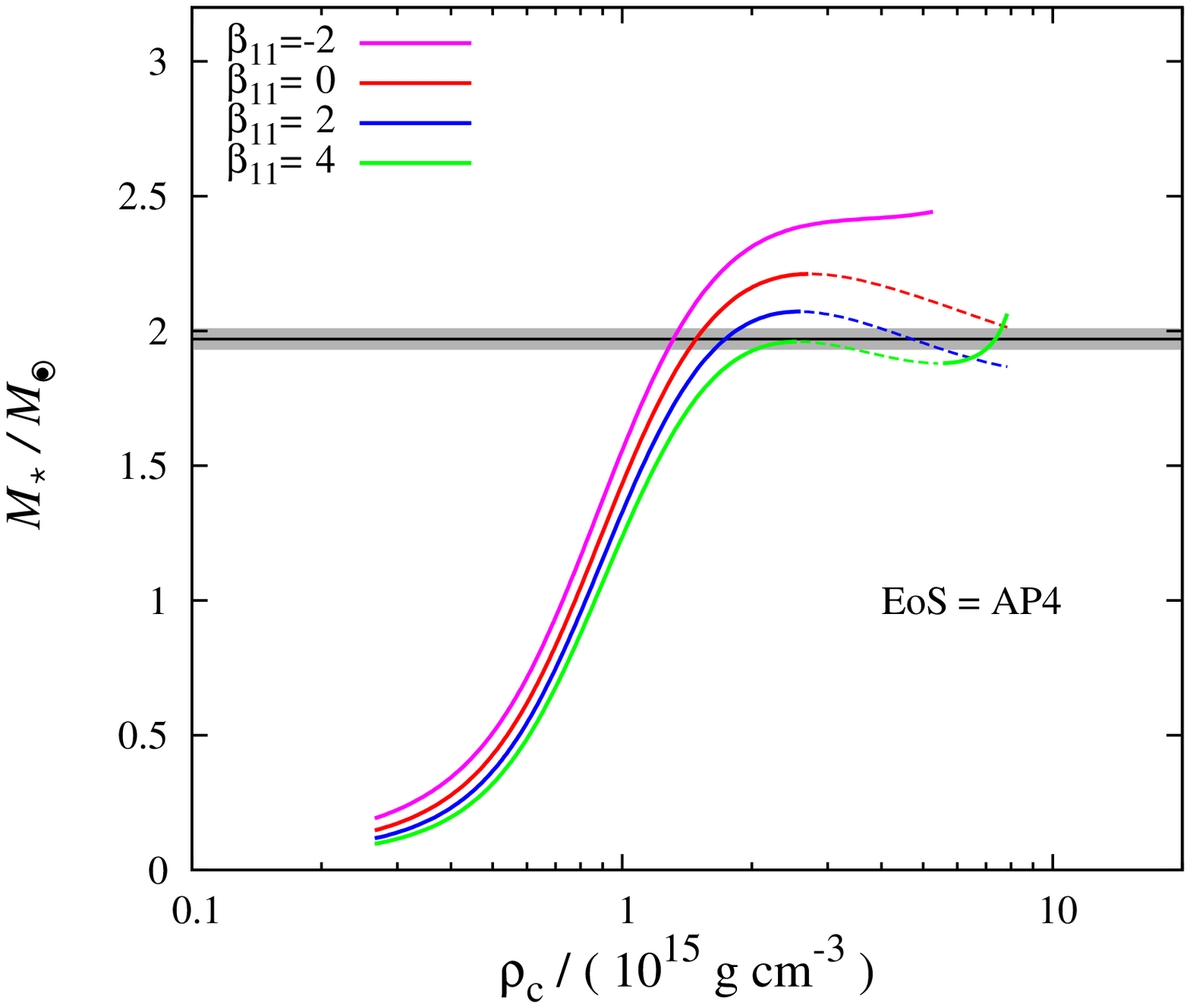}
\label{AP4_rho_M}
}\quad
\subfigure[]{
\includegraphics[width=0.45\textwidth]{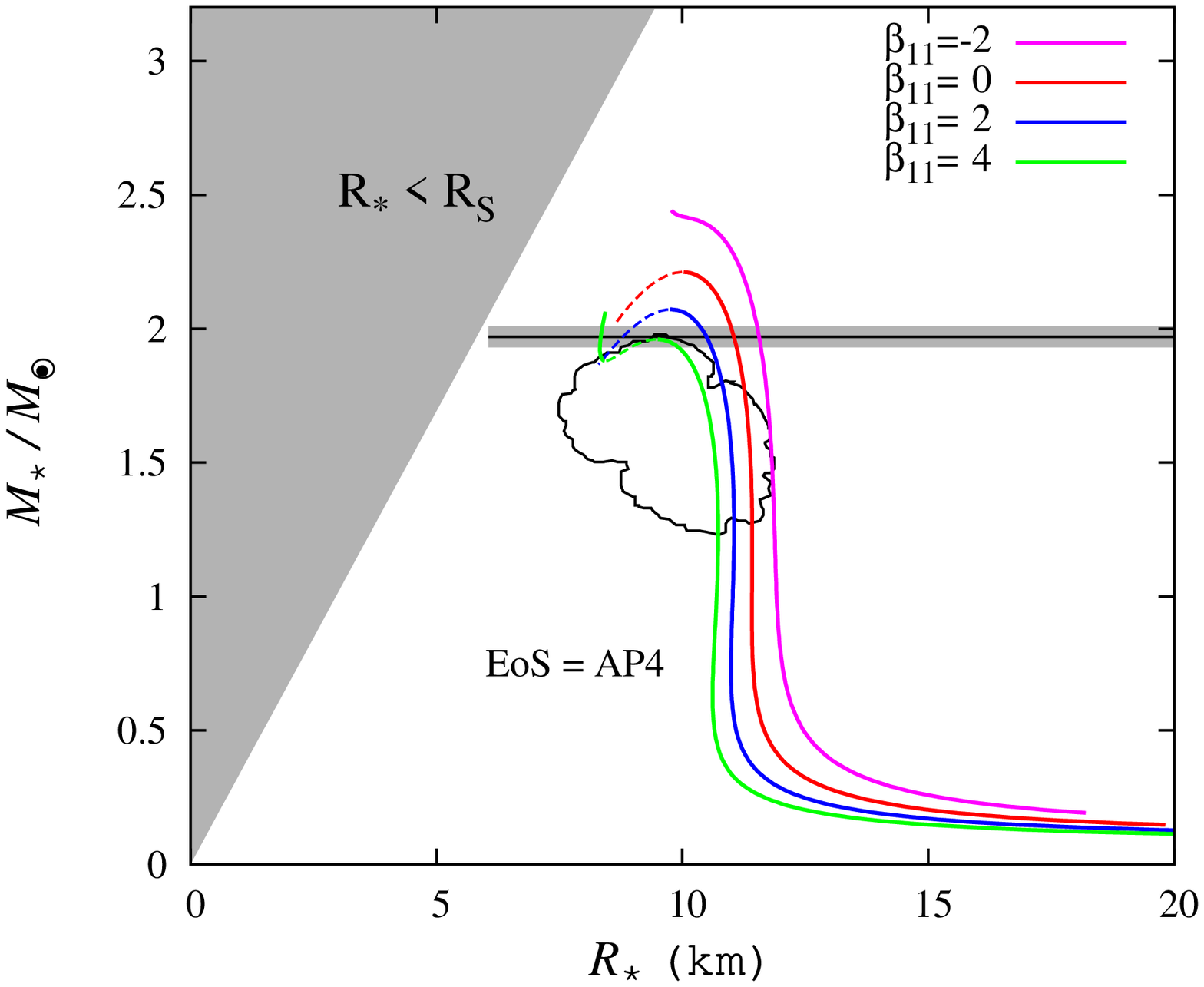}
\label{AP4_MR}
}
}
\centering
\caption{The $\rho_\mathrm{c}-M$ (panel a) and M-R relation (panel b) for AP4 EoS. See Figure~\ref{fig-FPS}
for the notation
in the figure. The results
are discussed in \S \ref{sub_sub_AP4}.
}
\label{fig-AP4}
\end{figure}

The results for AP4 are shown in Figure~\ref{fig-AP4}.
It is seen from the M-R relation that for $\beta_{11} >-2$, AP4 is consistent with the confidence contours of \cite{oze10}
while the maximum mass becomes lower than $\sim 2 M_{\odot}$ for $\beta_{11}>4$ and is not compatible with the existence of 
PSR J1614-2230 \cite{dem10}. Thus we conclude that $-2<\beta_{11}<4$ for AP4.

\subsubsection{GS1}
\label{sub_sub_GS1}
\begin{figure}
\centering
\mbox{
\subfigure[]{
\includegraphics[width=0.45\textwidth]{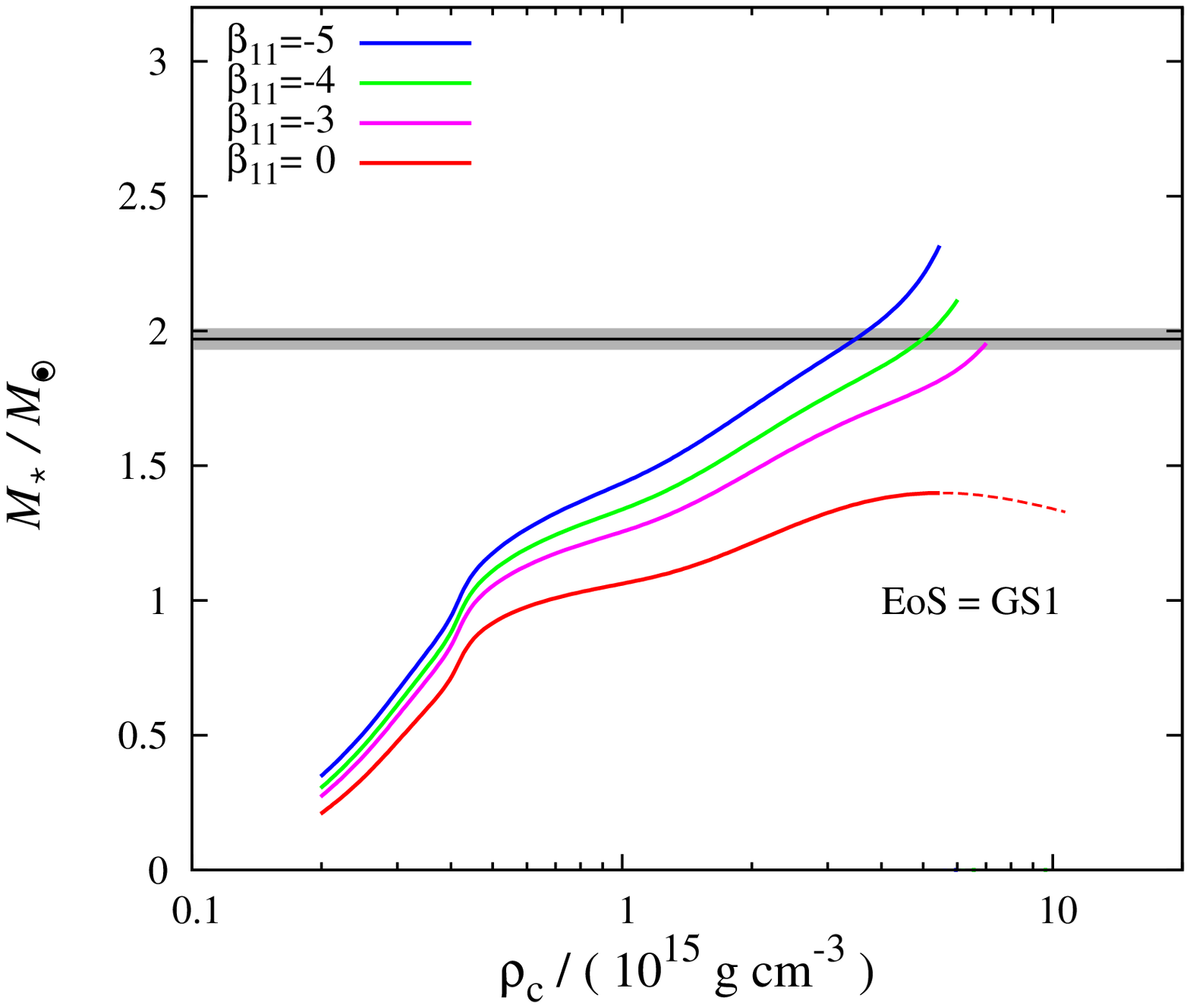}
\label{GS1_rho_M}
}\quad
\subfigure[]{
\includegraphics[width=0.45\textwidth]{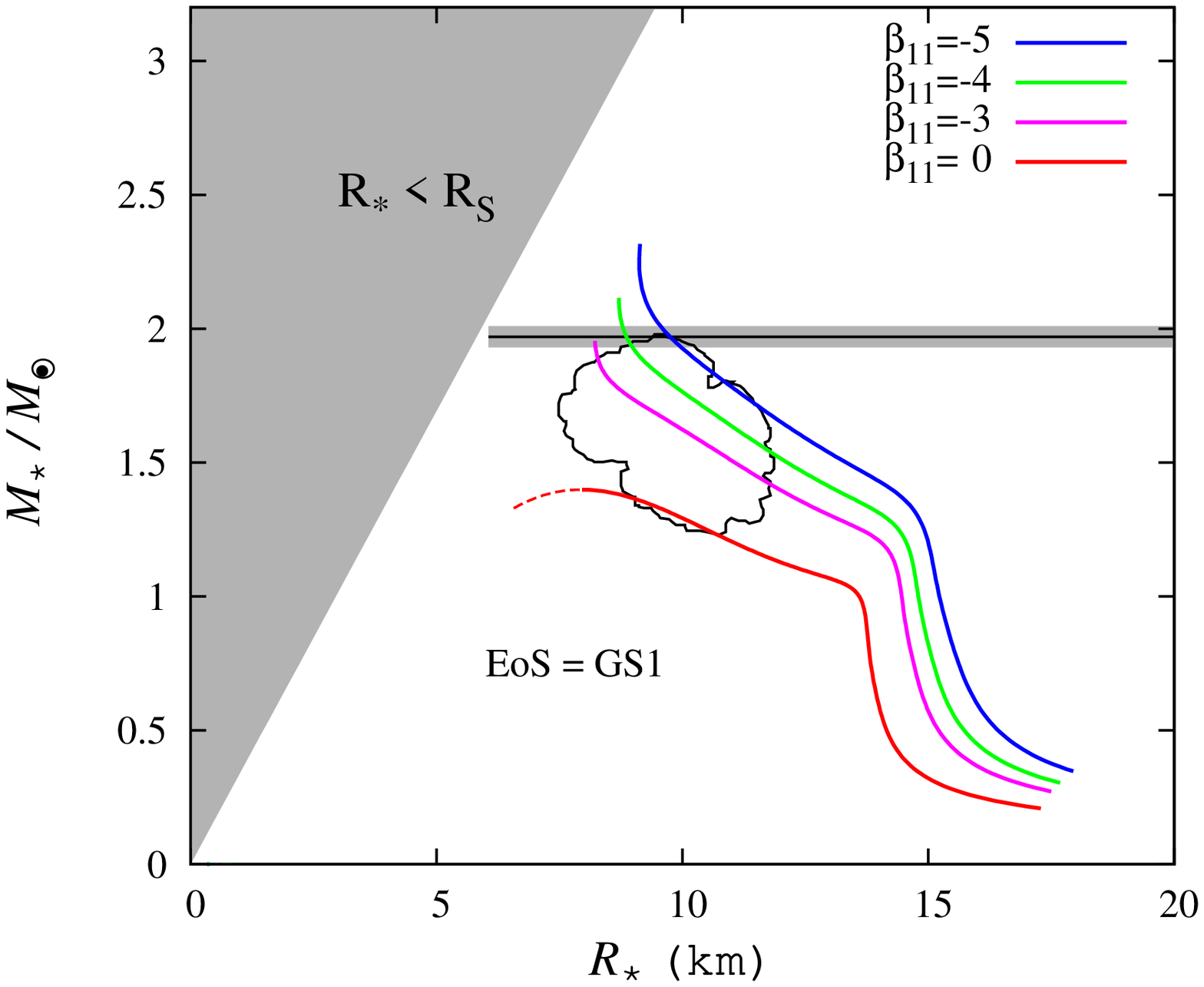}
\label{GS1_MR}
}
}
\centering
\caption{The $\rho_\mathrm{c}-M$ (panel a) and M-R relation (panel b) for GS1 EoS. 
See Figure~\ref{fig-FPS}
for the notation
in the figure. The results
are discussed in \S \ref{sub_sub_GS1}
}
\label{fig-GS1}
\end{figure}

The mass versus the central density, $\rho _{\mathrm{c}}-M$, and M-R relation for GS1 are shown 
in Figure~\ref{fig-GS1}.
GS1 is excluded within GR as its maximum mass is well below two solar masses.
The maximum mass reaches $\sim 2 \, M_{\odot}$ for $\beta_{11} < -3$. For $\beta_{11}<-5$, however, the M-R relation does not
pass through the confidence contours of \cite{oze10}. The observations than imply the constraint $-5<\beta_{11}<-3$ for GS1. 
  
For negative values of $\beta_{11}$ we find that the stability condition, $dM/d\rho _{\mathrm{c}}>0$, is satisfied for the whole range of central
densities considered. As no maximum mass is reached, this situation that does not have a counterpart in general relativity.

\subsubsection{MPA1}
\label{sub_sub_MPA1}
\begin{figure}
\centering
\mbox{
\subfigure[]{
\includegraphics[width=0.45\textwidth]{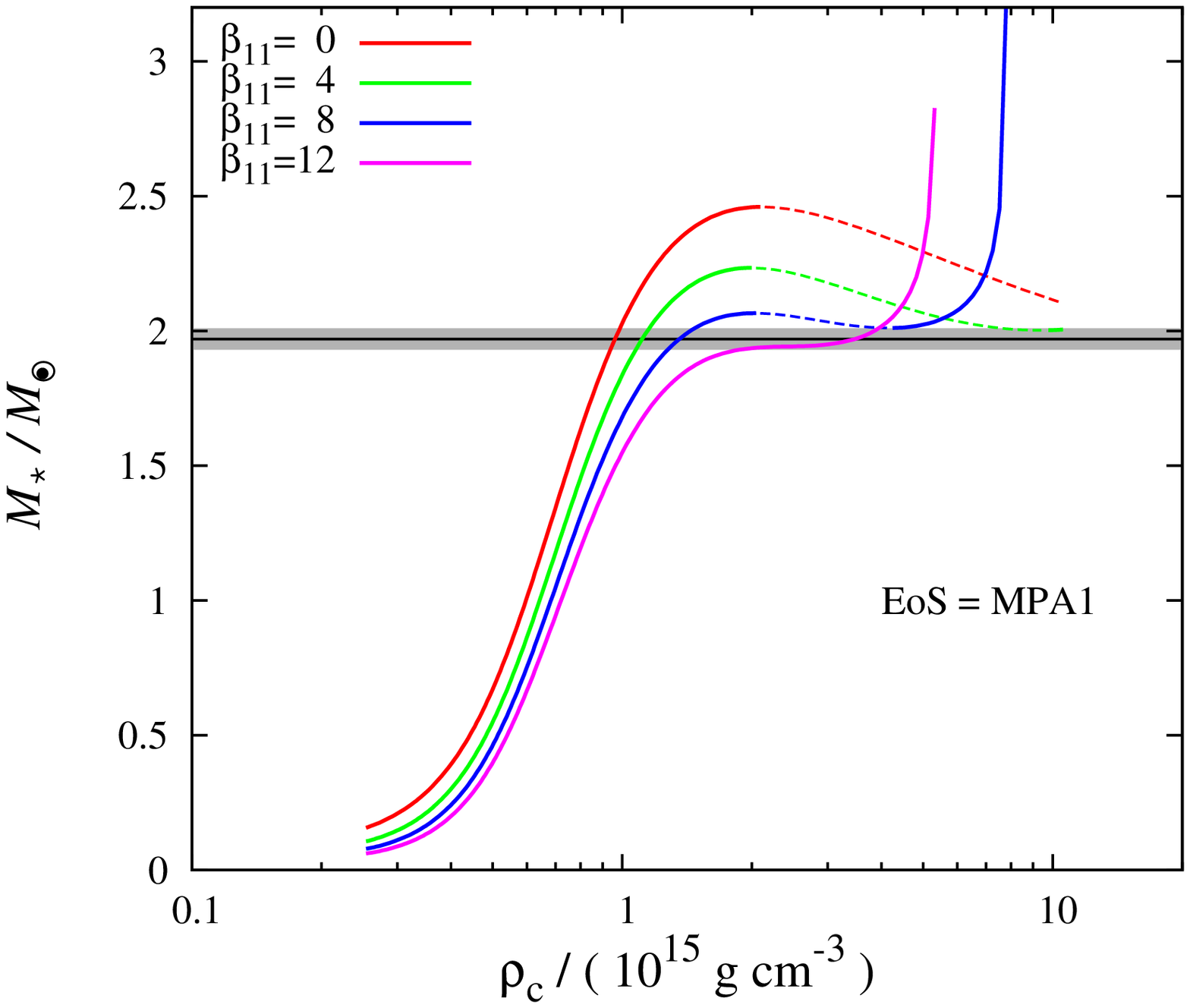}
\label{MPA1_rho_M}
}\quad
\subfigure[]{
\includegraphics[width=0.45\textwidth]{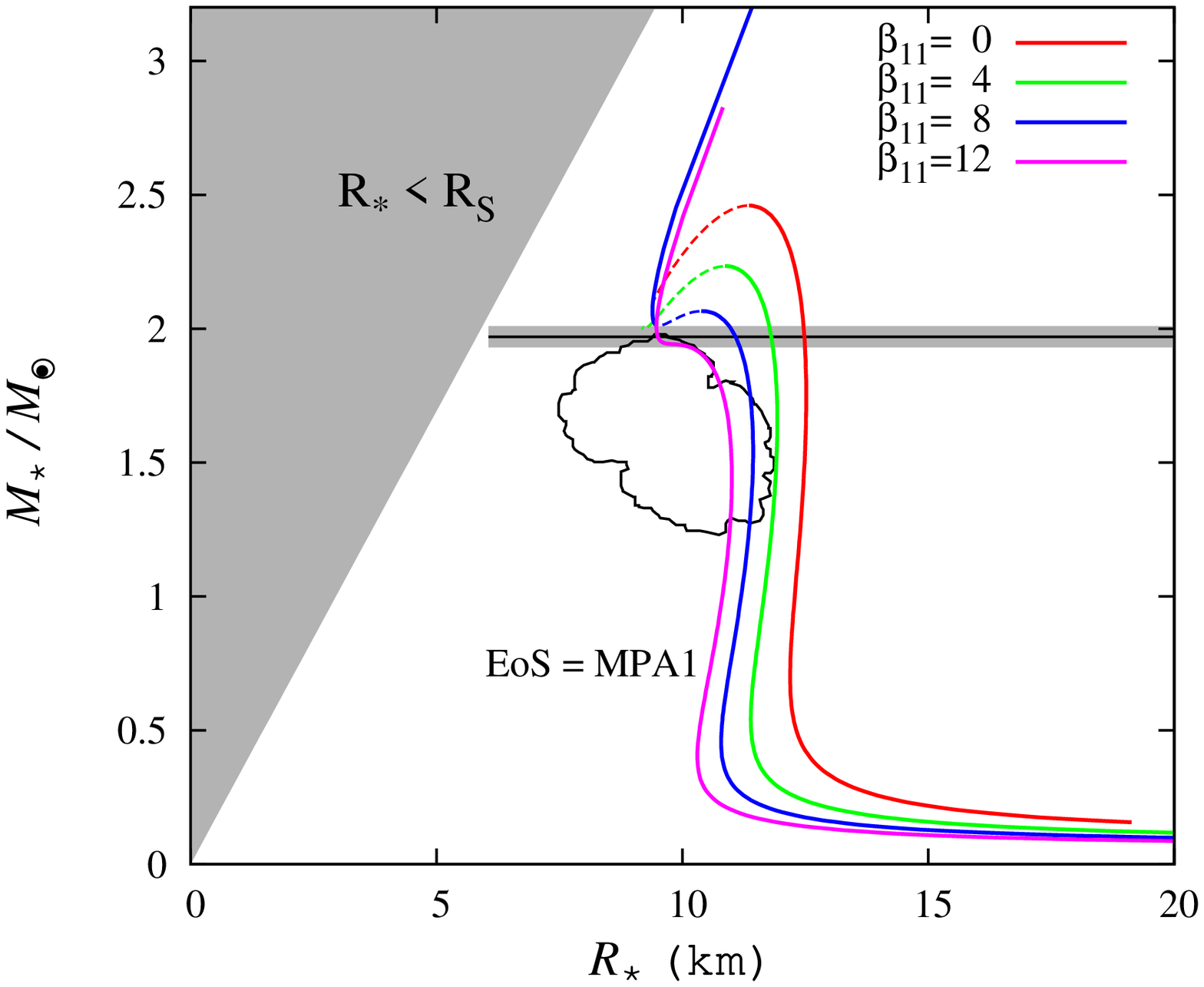}
\label{MPA1_MR}
}
}
\centering
\caption{The $\rho_\mathrm{c}-M$ (panel a) and M-R relation (panel b) for MPA1 EoS. 
See Figure~\ref{fig-FPS}
for the notation
in the figure. The results
are discussed in \S \ref{sub_sub_MPA1}
}
\label{fig-MPA1}
\end{figure}

The relation between the central density and the mass of the NS, $\rho _{\mathrm{c}}-M$, 
and M-R relation are shown in Figure~\ref{fig-MPA1}.
The maximum mass of MPA1 is above the observed mass of PSR J1614-2230 for $\beta_{11}<12$.
For $\beta_{11}<4$ the M-R relation does not pass through the confidence interval of \cite{oze10}. 
This then implies $4<\beta_{11}<12$ for MPA1. For larger values of $\beta$ the maximum mass 
obtained decreases as is the case for all EoSs. For $\beta_{11}=8$ we see that, apart from 
the usual stable (in the sense that $dM/d\rho_{\mathrm c}>0$) and unstable branches, 
there emerges a stable branch for the highest central pressures which does not have a 
counterpart in GR.
The unstable branch between the two stable branches disappears for even larger values 
of $\beta$. We thus observe that for 
$\beta_{11}=12$ we can obtain stable configurations with mass exceeding $2M_{\odot}$.

\subsubsection{MS1}
\label{sub_sub_MS1}
\begin{figure}
\centering
\mbox{
\subfigure[]{
\includegraphics[width=0.45\textwidth]{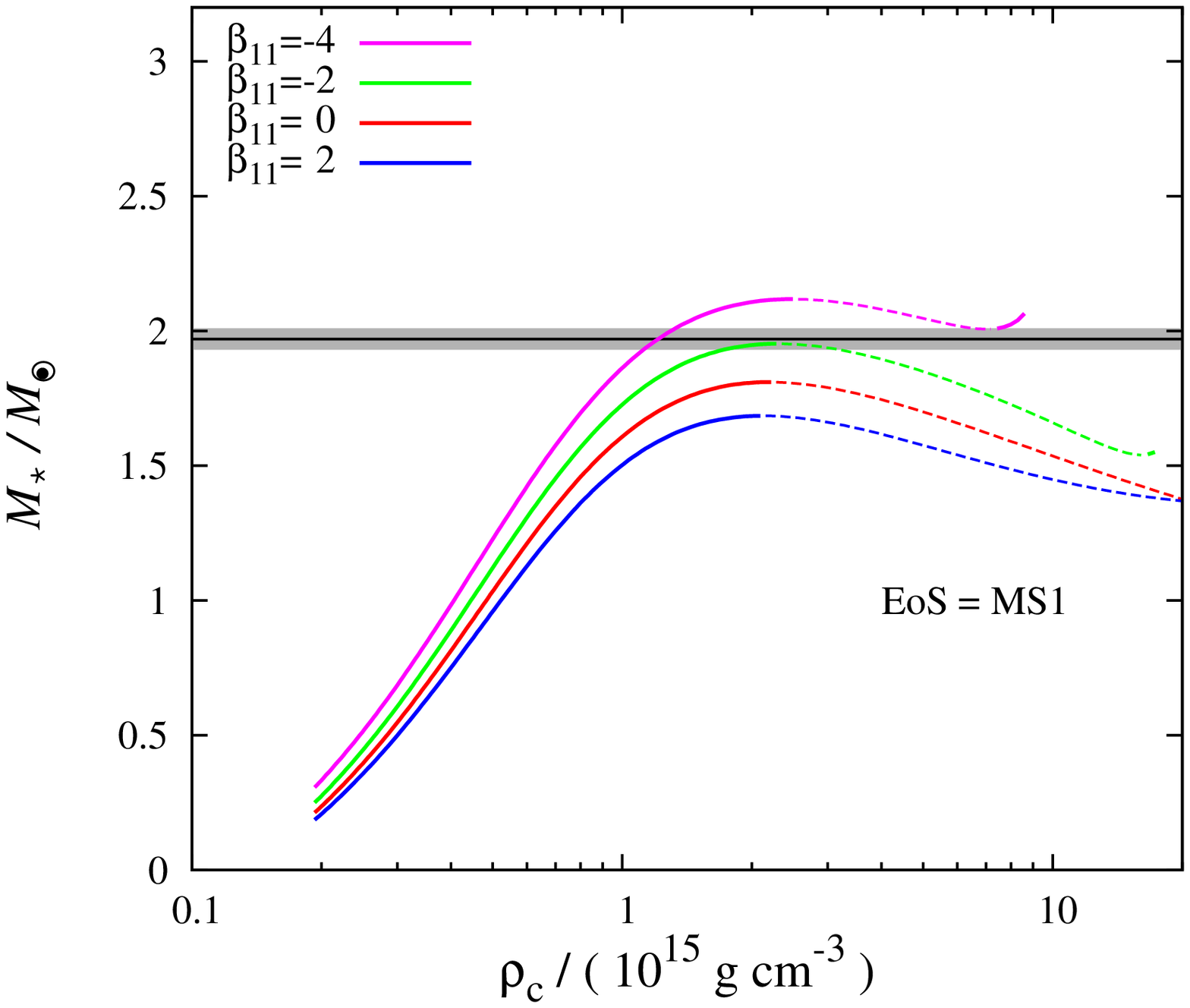}
\label{MS1_rho_M}
}\quad
\subfigure[]{
\includegraphics[width=0.45\textwidth]{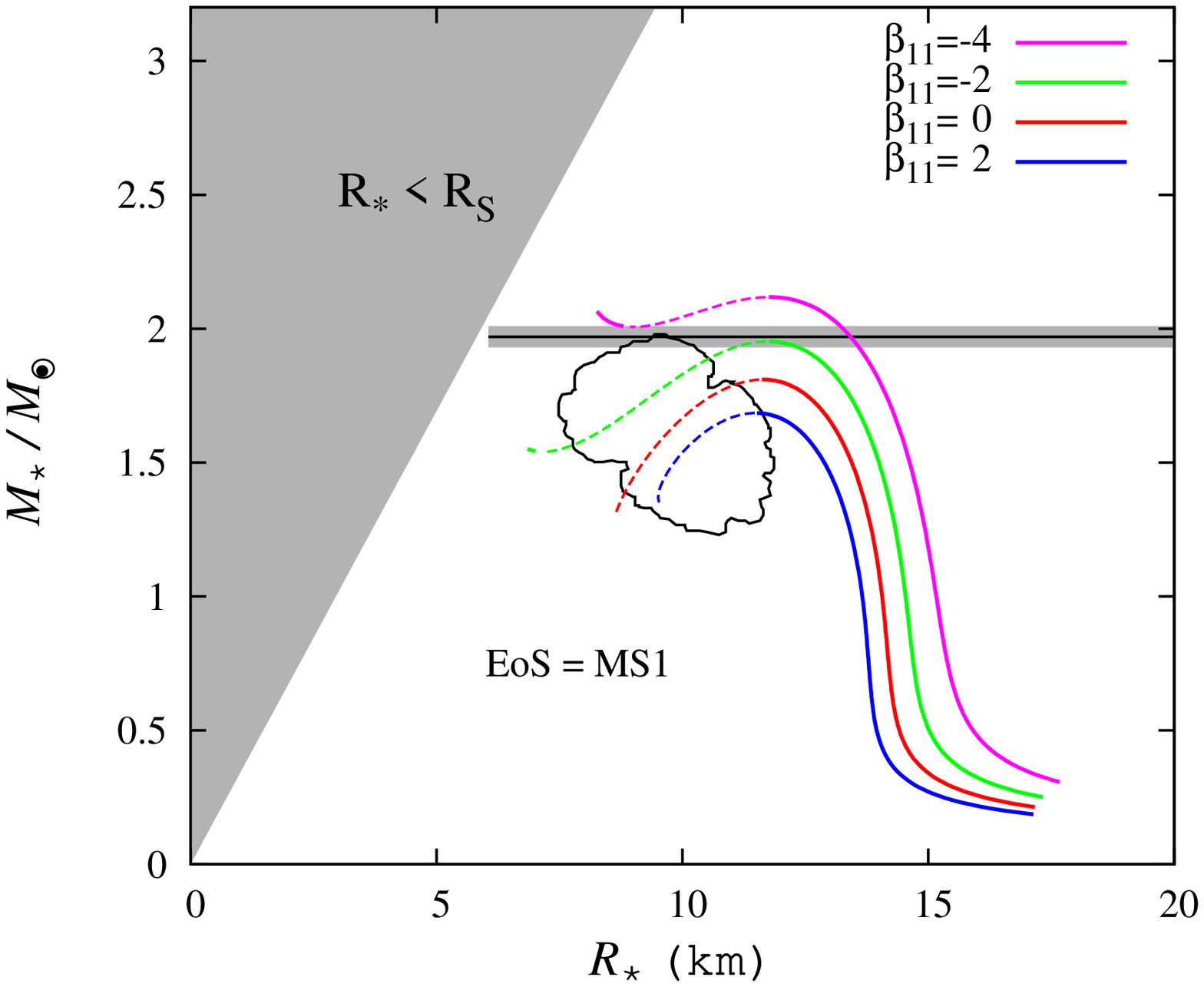}
\label{MS1_MR}
}
}
\centering
\caption{The $\rho_\mathrm{c}-M$ (panel a) and M-R relation (panel b(panel a)) for MS1 EoS. 
See Figure~\ref{fig-FPS}
for the notation
in the figure. The results
are discussed in \S \ref{sub_sub_MS1}
}
\label{fig-MS1}
\end{figure}

The $\rho_{\mathrm{c}}-M$ and M-R relation are shown in Figure~\ref{fig-MS1}.
This EoS has a maximum mass less than $2M_{\odot}$ in GR and does not
pass through the confidence contour presented in \cite{oze10}. In order that the M-R relation
has a maximum beyond the $2M_{\odot}$ measured mass one needs $\beta_{11}<-2$ which would take the M-R relation even further from the confidence contour of \cite{oze10}.
In order that the M-R relation marginally pass through the confidence contours of \cite{oze10} one needs $\beta_{11}>2$ for which the maximum mass obtained is even smaller. We conclude from this analysis that MS1 is excluded as a possible
EoS for NS in this gravity model assuming there is no systematic error in the M-R measurements of \cite{oze10}.

\subsection{Validity of the Perturbative Approach}

\begin{figure}[]
\begin{center}
\includegraphics[width=0.7\textwidth]{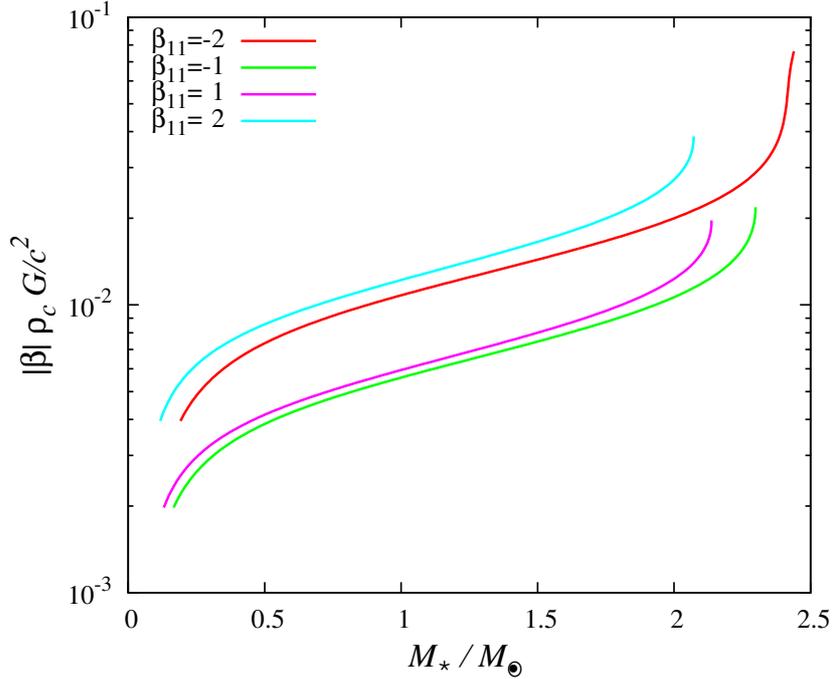}
\end{center}
\caption{The dependence of the dimensionless perturbation parameter 
$|\beta|\rho_{\mathrm c}G/c^2$ on mass. 
The result is obtained for the AP4 EoS. We see that the validity of the perturbative approach is questionable for large masses and large values of $\beta$. This leads us to conclude that the constraints we provide for each EoS may not be as accurate, since the dimensionless perturbation parameter reaching 0.1 signals the break-down of the validity of the perturbative approach.}
\label{fig-valid1}
\end{figure}

The validity of the perturbative method is justified if the terms that result from the $\beta R_{\mu \nu}R^{\mu \nu}$
term in the Lagrangian are smaller than the terms arising from $R$, the GR term. This requires that $|\beta|\rho_{\mathrm c} \ll c^2/G$ be satisfied for each configuration we consider. In Figure \ref{fig-valid1} we check this for EoS AP4 for a range of $\beta$ values.
We choose AP4 because it accommodates a large mass that could lead to the breakdown of the perturbative approach easily. We see that for large masses the dimensionless parameter that measures the perturbative term, $|\beta|\rho_{\mathrm c}G/c^2$, reaches
0.1 which is critical in deciding whether the perturbative approach is justified. This indicates that the constraints we provide
for $\beta$ may not be accurate for masses approaching $2M_{\odot}$ for $|\beta|\sim 10^{11}$ cm$^2$ for any EoS.

\begin{figure}[]
\begin{center}
\includegraphics[width=0.7\textwidth]{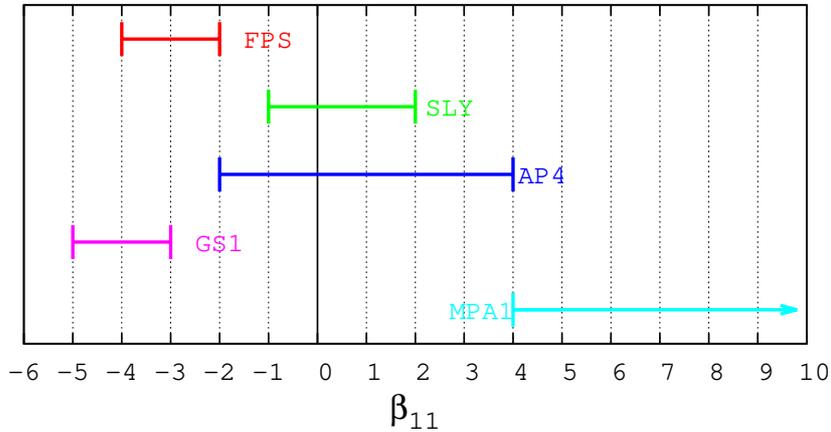}
\end{center}
\caption{The range of $\beta_{11}=\beta/(10^{11}\, \mathrm{cm}^2)$ consistent with the observations. MS1 is excluded from the plot, because it can not be reconciled with the confidence contours of \cite{oze10} for the gravity model considered though it predicts NS masses well above the measured mass of PSR J1614-2230 \cite{dem10} for $\beta_{11}<-2$. Note that $|\beta_{11}|>5$ 
is excluded for all EoSs we considered except for MPA1. The validity domain of MPA1 extends to very large positive values of $\beta$. We can not include its whole range as the perturbative approach applied in this work would break down for such large values.}
\label{beta_range}
\end{figure}

We observe from the figures in general that variations in the M-R relation 
comparable to employing different EoSs can be
obtained for $|\beta| \sim 10^{11}$ cm$^{2}$. Using $\beta \lesssim 10^{11}$
cm$^{2}$ gives M-R relations that can not be distinguished from the GR
results. The range of $\beta$ that is consistent with the observations are shown 
in Figure~\ref{beta_range} summarizing the results above. We see that $|\beta_{11}|>4$ 
is excluded for all EoSs we considered except for MPA1.

\subsection{Dependence of Maximum Mass on $\beta $}

For all EoSs we observe that the maximum stable mass of a neutron star, 
$M_{\max}$, and its radius at this mass, $R_{\min}$, increases for decreasing
values of $\beta$, for the ranges we consider in the figures. There is no
change in the behavior of $M_{\max}$ and $R_{\min}$ values as $\beta$
changes sign. Thus the structure of neutron stars in GR ($\beta=0$) does not
constitute an extremal configuration in terms of $M_{\max}$ and $R_{\min}$.

\begin{figure}[!]
\begin{center}
\includegraphics[width=0.8\textwidth]{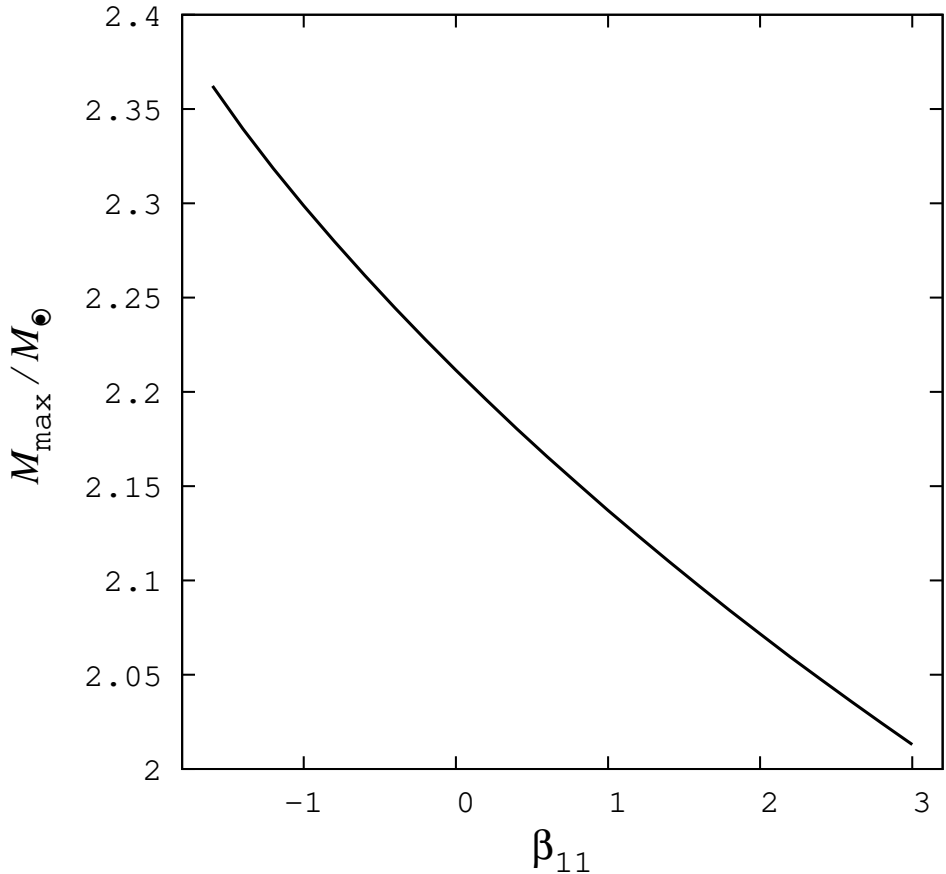}
\end{center}
\caption{$M_{\max}$ changing with $\beta$ for the AP4 EoS. Note that for $\beta_{11}<-1.8$, in this gravity model, there is no maximum mass but only stable configurations.}
\label{fig_beta}
\end{figure}

In Figure~\ref{fig_beta} we show the dependence of these quantities on the
value of $\beta$ for the EoS AP4. We see that $\beta =0$ is not special point in figure.  

\section{Conclusions}

In this paper we studied the structure of neutron stars (NSs) in a
generalized theory of gravity motivated by string theory. In the first
section we discussed the motivations for modifying gravity. In the second section we
derived the field equations of this gravity model from its action. In the
third section we obtained the hydrostatic equilibrium equations in spherical
symmetry from the field equations by using a perturbative approach in which
general relativity (GR) stands for the zeroth order gravity model. In the fourth section
we solved the hydrostatic equilibrium equations for NSs by using
numerical methods. In order to solve the equations we have used realistic
equations of state (EoSs) that describe the dense matter inside NSs and
obtained the mass-radius (M-R) relations depending on $\beta$, the free
parameter of the generalized gravity model considered. These M-R relations
are then compared with the recent observational measurements of mass and radius
of NSs to constrain the value of $\beta$.

We have shown that observationally significant changes on the M-R relation
are induced for $\beta \sim 10^{11}$ cm$^{2}$. An order of magnitude
smaller values for $\beta $ gives results that are not significantly
different from what GR predicts. An order of magnitude greater values, on
the other hand, leads to results that can not be associated with known
properties of NSs. For such values, the perturbative approach breaks
down too. For this selection of EoSs we see that none of them are consistent with the observations 
for $\beta<-5\times 10^{11}$ cm$^2$ and $\beta>4\times 10^{11}$ cm$^2$. The only exception is MPA1 which accepts
large positive values like  $\beta_{11}\sim 10$.
We thus conclude from this
analysis that $|\beta |\lesssim 5\times 10^{11}$ cm$^{2}$ is an upper limit brought
by observations of NSs \cite{oze10,dem10}.

We note that, for the the gravity model $R+\alpha R^2$, 
the constraint on $\alpha$ obtained by \cite{ara11}
by using NSs confronting the M-R relation with the same observations, is of the order $10^{10}$
cm$^2$, two orders of magnitude smaller than the constraint obtained on $\beta$ in the present work, 
although both $\alpha$ and $\beta$ have the same dimensions ($[L]^2$). This indicates that the $R^2$ 
term in eq. (\ref{rqugr}) is more effective on the structure of neutron stars than the $R_{\mu \nu }R^{\mu 
\nu }$ term.

The last observation is very interesting from a theoretical point of view. As mentioned in the 
introduction, theory described by (\ref{rqugr}) is equivalent to Weyl tensor squared modification of 
general relativity for $\beta = -3\alpha$ \cite{s08,fs09}. 
That is, from the point of view of Weyl tensor squared 
modification, $\alpha$ and $\beta$ are in the same order. Our perturbative analysis in this paper also in 
some way forces us to make a similar statement: if $\beta$ is taken in the same order as of $\alpha$, 
then there would not be the problem of the break-down of the perturbative approach. What extra we 
learn from this analysis is that even though the $R^2$ and $R_{\mu \nu }R^{\mu \nu }$ terms seem to 
be on equal footing in the expansion of a Weyl tensor square term,
\begin{equation}
\label{Weyl}
\frac12 C_{\mu \nu \rho \sigma}C^{\mu \nu \rho \sigma} = -\frac13 R^2 + R_{\mu \nu }R^{\mu \nu }
+ \mathrm{topological\, term},
\end{equation}
the contributions of them to neutron star physics are not equal. This is pointed out to us by the analysis 
in this paper and it is an important statement just by itself. It would be significant if this observation is 
confirmed and understood analytically. We plan to do this in a future publication.

We find that, some of the EoSs, which do not give M-R relations consistent
with the observations within the framework of GR, can be reconciled with
these observations via the free parameter $\beta $ in the generalized
gravity model considered in this work. This then brings up the question of
degeneracy between the EoSs and the free parameter $\beta $. This degeneracy
does not effect the constraint $|\beta|\lesssim 5\times 10^{11}$ cm$^{2}$ which is a bound
for all EoSs we considered except for MPA1.

We finally comment that the constraint we obtained is actually the strongest
constraint we could obtain by using NSs as the experimental apparatus.
Another estimate of the nominal value $\beta _{0}$ of the previous section
is as follows: Typical radius of a NS is $R_{\ast}\sim 10$ km, the only length
scale in the system. This corresponds to an estimate of curvature $R^{\mu
\nu }R_{\mu \nu }\sim R_{\ast}^{-2}\sim 10^{-12}$ cm$^{-2}$ and so the new
perturbative term $\beta R^{\mu \nu }R_{\mu \nu }$ will become of order $R_{\ast}$
and lead to variations on the structure of neutron stars for $\beta \sim
R_{\ast}^{2}\sim 10^{12}$ cm$^{2}$. As we obtain such variations in this limit, the 
value that we should obtain by using NSs, we
infer that the actual value of $\beta $ is likely much smaller than this. As
we mentioned before, deviations from GR are not significant for NSs for
values much less than this value.

\acknowledgments

This work is supported by the Turkish Council of Research and Technology (T%
\"{U}B\.{I}TAK) through grant number 108T686.


\begin{thebibliography}{99}

\bibitem{per99}
Supernova Cosmology Project Collaboration, S.~Perlmutter {\it et al.}, 
  \emph{Measurements of Omega and Lambda from 42 high redshift supernovae,
  Astrophys.\ J.\ }  {\bf 517} (1999) 565 [astro-ph/9812133].

\bibitem{rie98}
Supernova Search Team Collaboration, A.~G.~Riess {\it et al.},
  \emph{Observational evidence from supernovae for an accelerating universe and a cosmological 
  constant, Astron.\ J.\ }  {\bf 116} (1998) 1009 [astro-ph/9805201].

\bibitem{rie04}
Supernova Search Team Collaboration, A.~G.~Riess {\it et al.},
  \emph{Type Ia supernova discoveries at z > 1 from the Hubble Space Telescope:
  Evidence for past deceleration and constraints on dark energy evolution,
  Astrophys.\ J.\ }  {\bf 607} (2004) 665 [astro-ph/0402512].

\bibitem{Weinberg} 
  S.~Weinberg, \emph{The Cosmological Constant Problem,
  Rev.\ Mod.\ Phys.}\ \ {\bf 61} (1989) 1.
  
\bibitem{Peebles} 
  P.~J.~E.~Peebles and B.~Ratra, \emph{The Cosmological constant and dark energy,
  Rev.\ Mod.\ Phys.}\ \ {\bf 75} (2003) 559  [astro-ph/0207347].
  
\bibitem{Nobbenhuis} 
  S.~Nobbenhuis,
  \emph{Categorizing different approaches to the cosmological constant problem,
  Found.\ Phys.}\ \ {\bf 36} (2006) 613 [gr-qc/0411093].
  
\bibitem{Bousso} 
  R.~Bousso, \emph{TASI Lectures on the Cosmological Constant,
  Gen.\ Rel.\ Grav.}\ \ {\bf 40} (2008) 607 [arXiv:0708.4231].

\bibitem{car01}
  S.~M.~Carroll, \emph{The Cosmological Constant,
  Living Rev.\ Rel.\ } {\bf 4} (2001) 1 [astro-ph/0004075].

\bibitem{uzan}
  J.~P.~Uzan, \emph{The acceleration of the universe and the physics behind it,
  Gen.\ Rel.\ Grav.\ }  {\bf 39} (2007) 307 [astro-ph/0605313].

\bibitem{tsu10}
  S.~Tsujikawa, \emph{Modified gravity models of dark energy,
  Lect.\ Notes Phys.\ } {\bf 800}  (2010) 99 [arXiv:1101.0191].

\bibitem{man05} 
  P.~D.~Mannheim, \emph{Alternatives to dark matter and dark energy,
  Prog.\ Part.\ Nucl.\ Phys.\ }  {\bf 56} (2006) 340 [astro-ph/0505266].
  
\bibitem{clif11} 
  T.~Clifton, P.~G.~Ferreira, A.~Padilla and C.~Skordis,
  \emph{Modified Gravity and Cosmology},  arXiv:1106.2476.

\bibitem{psa08}
  D.~Psaltis, \emph{Probes and Tests of Strong-Field Gravity with Observations
  in the Electromagnetic Spectrum,} arXiv:0806.1531.

\bibitem{Odintsov-rev} 
  S.~Nojiri and S.~D.~Odintsov,
  \emph{Introduction to modified gravity and gravitational alternative for dark energy,
  Int.\ J.\ Geom.\ Meth.\ Mod.\ Phys.}\ \ {\bf 4} (2007) 115 [hep-th/0601213].
  
\bibitem{Capozziello-rev} 
  S.~Capozziello and M.~Francaviglia,
  \emph{Extended Theories of Gravity and their Cosmological and Astrophysical Applications,
  Gen.\ Rel.\ Grav.}\ \ {\bf 40} (2008) 357 [arXiv:0706.1146].
  
\bibitem{Sotiriou-rev} 
  T.~P.~Sotiriou and V.~Faraoni, \emph{f(R) Theories Of Gravity,
  Rev.\ Mod.\ Phys.}\ \ {\bf 82} (2010) 451 [arXiv:0805.1726].
  
\bibitem{deFelice-rev} 
  A.~De Felice and S.~Tsujikawa, \emph{f(R) theories,
  Living Rev.\ Rel.}\ \ {\bf 13} (2010) 3 [arXiv:1002.4928].

\bibitem{ara11}
  A.~S.~Arapoglu, C.~Deliduman, and K.~Yavuz Eksi,
  \emph{Constraints on Perturbative f(R) Gravity via Neutron Stars,
  JCAP} {\bf 1107} (2011) 020 [arXiv:1003.3179].
  
\bibitem{mag94} 
  G.~Magnano and L.~M.~Sokolowski, \emph{On physical equivalence between nonlinear gravity 
  theories and a general relativistic selfgravitating scalar field, Phys.\ Rev.} {\bf D\ 50} (1994) 5039
  [gr-qc/9312008].

\bibitem{bd} 
  C.~Brans and R.~H.~Dicke, \emph{Mach's principle and a relativistic theory of gravitation,
  Phys.\ Rev.}\ \ {\bf 124} (1961) 925.
  
\bibitem{zwi85} 
  B.~Zwiebach, \emph{Curvature Squared Terms and String Theories,
  Phys.\ Lett.\ } {\bf B 156} (1985) 315.
  
\bibitem{psa09} 
  D.~Psaltis, \emph{Two approaches to testing general relativity in the strong-field regime,
  J.\ Phys.\ Conf.\ Ser.} {\bf 189} (2009) 012033 [arXiv:0907.2746].
  
\bibitem{par99} 
  L.~Parker and A.~Raval,
  \emph{Nonperturbative effects of vacuum energy on the recent expansion of the universe,
   Phys.\ Rev.\ } {\bf D 60} (1999) 063512 [gr-gc/9905031].

\bibitem{s08}
  I.~L.~Shapiro,
  \emph{Effective Action of Vacuum: Semiclassical Approach,
  Class.\ Quant.\ Grav.\ }  {\bf 25}, 103001 (2008) [arXiv:0801.0216].  

\bibitem{fs09}  
V.~P.~Frolov and I.~L.~Shapiro,
  \emph{Black Holes in Higher Dimensional Gravity Theory with Quadratic in Curvature Corrections,
  Phys.\ Rev.\ } {\bf D 80}, 044034 (2009) [arXiv:0907.1411].
  
\bibitem{bay02} 
  S.~Deser and B.~Tekin,
  \emph{Gravitational energy in quadratic curvature gravities,
  Phys.\ Rev.\ Lett.\ } {\bf 89} (2002) 101101 [hep-th/0205318].

\bibitem{emi10}
  E.~Santos,
  \emph{Quantum vacuum effects as generalized f(R) gravity. Application to stars, 
  Phys.\ Rev.\ }  {\bf D 81} (2010) 064030 [arXiv:0909.0120].

\bibitem{coo10}
  A.~Cooney, S.~Dedeo, and D.~Psaltis, 
 \emph{Neutron Stars in f(R) Gravity with Perturbative Constraints,
  Phys.\ Rev.\ } {\bf D 82} (2010) 064033 [arXiv:0910.5480].

\bibitem{emi11}
  E.~{Santos}, \emph{Neutron stars in generalized f(R) gravity, Astrop.\ \& Space Sci.\ accepted},  [arXiv:1104.2140].
  
\bibitem{oze10}
  F.~{\"O}zel, G.~Baym, and T.~G{\"u}ver,
  \emph{Astrophysical Measurement of the Equation of State of Neutron Star Matter,
  Phys.\ Rev.\ } {\bf D 82} (2010) 101301 [arXiv:1002.3153].

\bibitem{dem10}
  P.~Demorest, T.~Pennucci, S.~Ransom, M.~Roberts and J.~Hessels,
  \emph{Shapiro Delay Measurement of A Two Solar Mass Neutron Star,
  Nature }  {\bf 467} (2010) 1081 [arXiv:1010.5788].

\bibitem{tol39}
  R.~C.~Tolman,
  \emph {Static solutions of Einstein's field equations for spheres of fluid,
  Phys.\ Rev.\ } {\bf 55} (1939) 364.

\bibitem{opp39}
  J.~R.~Oppenheimer and G.~M.~Volkoff,
  \emph{On Massive neutron cores, 
   Phys.\ Rev.\ } {\bf 55} (1939) 374.

\bibitem{ja86}
   X.~Jaen, J.~Llosa and A.~Molina,
  \emph{A Reduction of order two for infinite order lagrangians,
  Phys.\ Rev.\ } {\bf D 34} (1986) 2302.

\bibitem{eli89}
  D.~A.~Eliezer and R.~P.~Woodard,
  \emph{The Problem of Nonlocality in String Theory,
  Nucl.\ Phys.\ } {\bf B 325} (1989) 389.

\bibitem{ded08}
  S.~DeDeo and D.~Psaltis,
  \emph{Stable, accelerating universes in modified-gravity theories,
  Phys.\ Rev.\ } {\bf D 78} (2008) 064013 [arXiv:0712.3939].

\bibitem{coo09}
  A.~Cooney, S.~Dedeo, and D.~Psaltis,
  \emph{Gravity with Perturbative Constraints: Dark Energy Without New Degrees of Freedom,
  Phys.\ Rev.\ } {\bf D 79} (2009) 044033 [arXiv:0811.3635].

\bibitem{ge11}
  C.~G{\"u}ng{\"o}r and K.~Y.~Eksi,
  \emph{Analytical Representation for Equations of State of Dense Matter,}
 [arXiv:1108.2166].

\bibitem{ref_FPS}
  V.R. Pandharipande and D. G. Ravenhall,
  \emph{Hot Nuclear Matter, in Nuclear Matter and Heavy Ion Collisions,
  NATO ADS Ser.} {\bf B 205} (1989) 103.

\bibitem{ref_AP4}
  A.~Akmal and V.~R.~Pandharipande,
  \emph{Spin-isospin structure and pion condensation in nucleon matter,
  Phys.\ Rev.\ } {\bf 56}  (1997) 2261 [nucl-th/9705013].
  
\bibitem{ref_SLY}
  F.~Douchin and P.~Haensel,
  \emph{A Unified Equation of State of Dense Matter and Neutron Star Structure,
  Astron. Astrophys. } \textbf{380} (2001) 151 [astro-ph/0111092].

\bibitem{ref_MS1}
  H.~M{\"u}ller and B.D.~Serot,
  \emph{Relativistic mean-field theory and the high-density nuclear equation of state,
  Nucl.\ Phys.\ } {\bf A 606} (1996) 508 [nucl-th/9603037].

\bibitem{ref_MPA1}
  H.~M{\"u}ther, M.~Prakash, and T.~L.~Ainsworth,
  \emph{The nuclear symmetry energy in relativistic Brueckner-Hartree-Fock calculations,
  Physics Letters } {\bf B 199} (1987) 469.

\bibitem{ref_GS1}
  N.~K.~Glendenning and J.~Schaffner-Bielich,
  \emph{First order kaon condensate,
  Phys.\ Rev.\ }  {\bf C 60} (1999) 025803 [astro-ph/9810290].

\bibitem{lat01}
  J.~M.~Lattimer and M.~Prakash,
  \emph{Neutron star structure and the equation of state,
  Astrophys.\ J.\ }  {\bf 550} (2001) 426 [astro-ph/0002232].

\bibitem{oze09}
  F.~{\"O}zel and D.~Psaltis,
  \emph{Reconstructing the Neutron-Star Equation of State from Astrophysical Measurements,
  Phys.\ Rev.\ }  {\bf D 80}  (2009) 103003 [arXiv:0905.1959].

\bibitem{oze1745}
  F.~{\"O}zel, T.~G{\"u}ver, and D.~Psaltis,
  \emph{The Mass and Radius of the Neutron Star in EXO 1745-248,
  Astrophys.\ J.\ }  {\bf 693} (2009) 1775 [arXiv:0810.1521].

\bibitem{guv1608}
  T.~G{\"u}ver, F.~{\"O}zel, A.~Cabrera-Lavers, and P.~Wroblewski, 
  \emph{The Distance, Mass, and Radius of the Neutron Star in 4U 1608-52,
  Astrophys.\ J.\ }  {\bf 712} (2010) 964 [arXiv:0811.3979].

\bibitem{guv1820}
  T.~G{\"u}ver, P.~Wroblewski, L.~Camarota, and F.~{\"O}zel, 
   \emph{The Mass and Radius of the Neutron Star in 4U 1820-30,
  Astrophys.\ J.\ }  {\bf 719} (2010) 1807 [arXiv:1002.3825].

\bibitem{ste10}
A.~W.~Steiner, J.~M.~Lattimer, and E.~F.~Brown, \emph{The Equation of
  State from Observed Masses and Radii of Neutron Stars, Astrophys.\ J.\ } {\bf
  722} (2010) 33 [arXiv:1005.0811].
  
\end{thebibliography}
\end{document}